\documentclass[sigconf, review=false, anonymous=false]{acmart}

\usepackage{subcaption}
\usepackage{enumitem}
\usepackage{listings}
\usepackage{soul,color}
\usepackage{subcaption}
\usepackage{hyperref}
\usepackage{soul}
\usepackage{algorithmic}
\usepackage{graphicx}
\usepackage{textcomp}

\hypersetup{
  hidelinks,
  colorlinks=true,
  allcolors=black,
  pdfstartview=Fit,
  breaklinks=true
}

\AtBeginDocument{%
  \providecommand\BibTeX{{%
    \normalfont B\kern-0.5em{\scshape i\kern-0.25em b}\kern-0.8em\TeX}}}

\setcopyright{none}
\renewcommand\footnotetextcopyrightpermission[1]{}

\usepackage{tcolorbox}
\usepackage{fontawesome5}
\usepackage{makecell}
\usepackage[framemethod=TikZ]{mdframed}
\usepackage{xcolor}
\usepackage{flushend}

\begin{document}

\title{Toward the Cure of Privacy Policy Reading Phobia: Automated Generation of Privacy Nutrition Labels From Privacy Policies}

\author{Shidong Pan\footnotemark[1]\footnotemark[2], Thong Hoang\footnotemark[2], Dawen Zhang\footnotemark[1]\footnotemark[2], Zhenchang Xing\footnotemark[1]\footnotemark[2], Xiwei Xu\footnotemark[2], Qinghua Lu\footnotemark[2], and Mark Staples\footnotemark[2]}
\vspace{5pt}
\affiliation{
  \institution{\footnotemark[1]School of Computing, Australian National University}
  \country{Canberra, Australia}
}
\email{{Shidong.Pan, Zhenchang.Xing}@anu.edu.au}
\affiliation{
  \institution{\footnotemark[2]Software Systems Research Group, CSIRO's Data61, Australia}
  \country{Sydney, Australia}
}
\vspace{5pt}
\email{{James.Hoang, Dawen.Zhang, Xiwei.Xu, Qinghua.Lu, Mark.Staples}@data61.csiro.au}

\renewcommand{\shortauthors}{Pan et al.}

\begin{abstract}

Software applications have become an omnipresent part of modern society. The consequent privacy policies of these applications play a significant role in informing customers how their personal information is collected, stored, and used. However, customers rarely read and often fail to understand privacy policies because of the ``Privacy Policy Reading Phobia'' \textit{(PPRP)}.
To tackle this emerging challenge, we propose the first framework that can automatically generate privacy nutrition labels from privacy policies. 
Based on our ground truth applications about the Data Safety Report from the Google Play app store, our framework achieves a 0.75 F1-score on generating first-party data collection practices and an average of 0.93 F1-score on general security practices.
We also analyse the inconsistencies between ground truth and curated privacy nutrition labels on the market, and our framework can detect 90.1\% under-claim issues.
Our framework demonstrates decent generalizability across different privacy nutrition label formats, such as Google's Data Safety Report and Apple's App Privacy Details.
\end{abstract}

\keywords{Privacy Nutrition Labels, Privacy Policy, Large Language Model}

\maketitle
\pagestyle{plain}

\section{Introduction}
\label{sec_introduction}

Software applications have become an omnipresent part of modern society. These applications collect and use a tremendous amount of personal information from customers to enhance user experiences.
However, the collected data is often misused for undisclosed actions, such as targeted advertising~\cite{ullah2020privacy}, price discrimination~\cite{mikians2012detecting}, or gender discrimination~\cite{datta2014automated}, which inevitably raise privacy concerns among the customers. Privacy policies are one of the most common methods to inform customers how their personal information is collected, stored, and used~\cite{caramujo2015analyzing, yu2015autoppg, perez2018review, kemp2020concealed}. Hence, privacy policies are a vital component of responsible technology in any software application ecosystems.

\begin{figure}[htbp]
\centering
\includegraphics[width=1.0\linewidth]{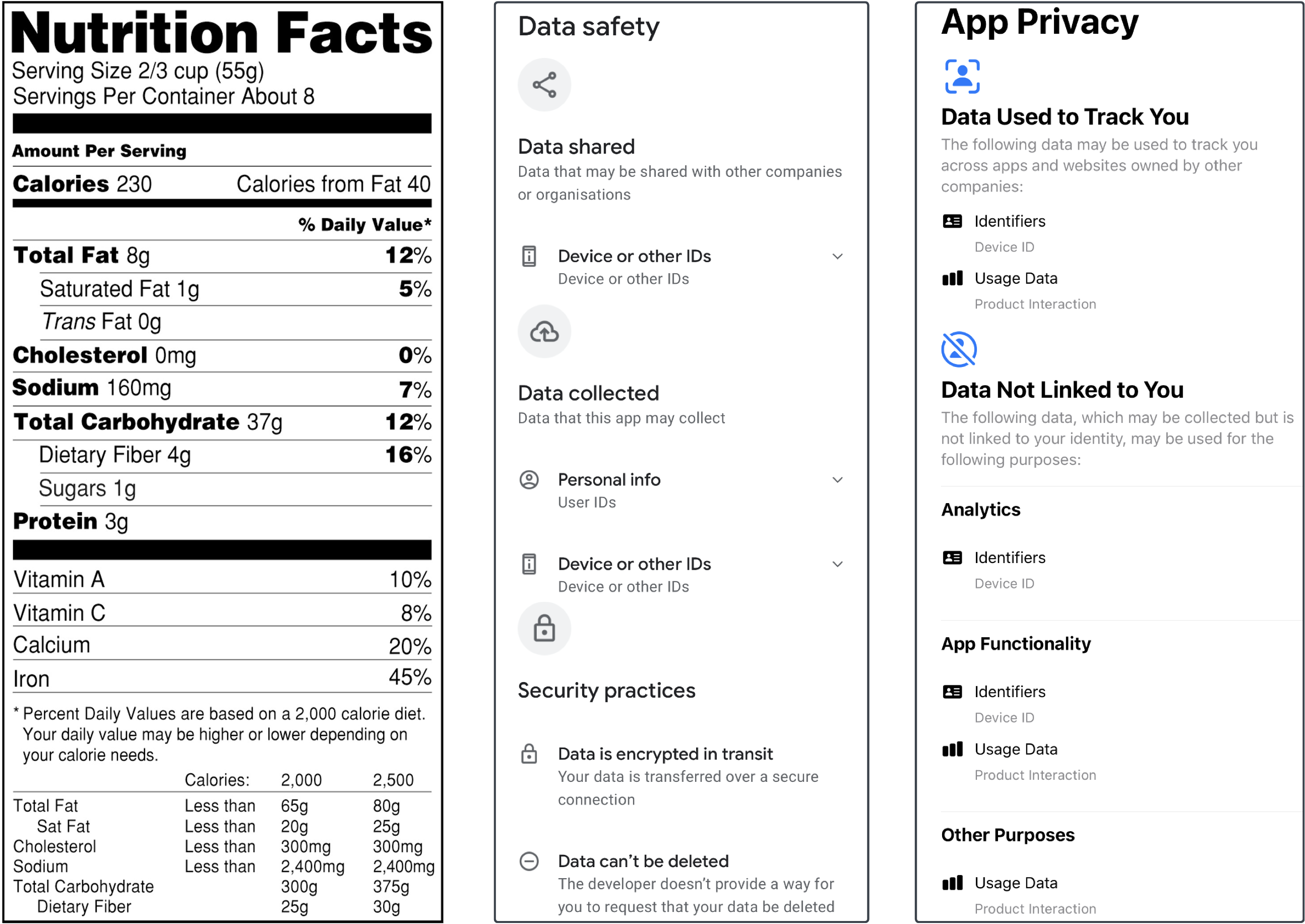}
\caption{
The left image is a nutrition facts label format published by the U.S. Food \& Drug Administration in 2016~\cite{foodlabel}. The middle and right images are screenshots of privacy nutrition labels from the Google Play app store~\cite{googledatasafety} and Apple App Store~\cite{appleprivacy}, respectively.
}
\label{fig_privacylabel}
\end{figure}

However, customers rarely read and often fail to understand privacy policies~\cite{obar2020biggest}. The privacy policies of software applications are often written in lengthy descriptions and overwhelming details, including technical and legal terms, leaving customers unable to comprehend their content.  According to a recent survey conducted by The Washington Times\footnote{https://www.washingtonpost.com/technology/2022/05/31/abolish-privacy-policies/}, 36\% of interviewees have \textbf{never} read privacy policies, and only 9\% have \textbf{always} read privacy policies before agreeing to a software application's terms. We call this specific quasi-bibliophobia as ``Privacy Policy Reading Phobia'' (\textit{PPRP}). The three main reasons causing the \textit{PPRP} problem are presented as follows:

\begin{itemize} [leftmargin=*]
    \item 
    Privacy policies require their aspects, such as legal, technical, and social, to be explained clearly and understandably~\cite{Blakkarly2022privacy}. 
    As the legal and technical statements need specific domain knowledge, the software application providers may employ explicit expert language to describe their information. 
    As a result, customers who lack technical or legal education may struggle to understand privacy policies. Some research studies found that most privacy policies require at least a college-level education to comprehend their content~\cite{kincaid1975derivation, krumay2020readability}, leading customers  
    to feel uneasy about spending time reading and understanding the privacy policies. 

    \item Privacy policies often provide lengthy and detailed descriptions
    of data practices in software applications, leading to information overload~\cite{krumay2020readability, Blakkarly2022privacy}. 
    A recent study shows that the privacy policies of 75 leading applications and websites contain an average of 4,000 words and require approximately 16 minutes to read completely~\cite{Blakkarly2022privacy}. 
    Moreover, the average reading time for a whole privacy policy is only 73 seconds~\cite{obar2020biggest}; hence, it will be challenging for customers to understand the complete content of the privacy policy.

    \item The ``Privacy Paradox'' refers to the phenomenon where individuals express concerns about their privacy but, in practice, engage in behaviors that compromise their own privacy~\cite{barth2017privacy}. 
    It is commonly attributed to that customers often have a confirmation bias~\cite{nickerson1998confirmation, mercier2016confirmation, lewicka1998confirmation}.
    Specifically, as customers often lack awareness of privacy protection laws, they tend to accept the software applications' privacy policies because they believe that these applications are harmless~\cite{dommeyer2003consumers, xie2019consumers, tinggi2011customers}. 
    
\end{itemize}

As the key is that software applications are used by broad groups of people in life and work, there is an urgent need to provide a more concise and transparent form of privacy policies, thus helping customers to save time reading and understanding these policies~\cite{mulder2019health, mohan2019analyzing, tang2021defining}. 
Concise, transparent and understandable language to deliver privacy information are also commonly required in privacy regulations such as GDPR~\cite{GDPR} [Art. 12 (1)] and CCPA~\cite{CCPA} [Regs \textsection{}999.308(a)(2)(d)]. 
Researchers have proposed \textbf{\textit{privacy nutrition labels}}~\cite{kelley2009nutrition, kelley2010standardizing, kelley2013privacy}
to offer a structured and succinct way of informing customers about how their information is collected and used in practice. 
Kelly et al.~\cite{kelley2009nutrition} defined privacy nutrition labels aiming to improve the visual presentation and comprehensibility of privacy policies for customers to mitigate the problem of lengthy and detailed policies. Specifically, they designed privacy nutrition labels by employing a two-dimensional grid layout with titles and bold texts to reduce the complexity of privacy policies. Following this work, Kelly et al.~\cite{kelley2010standardizing} also found that standardized privacy nutrition labels may increase the speed of comprehension and enjoyment of reading for customers. Later on, some research work focuses on designing privacy nutrition labels for mobile applications~\cite{kelley2013privacy} and Internet of Things (IoT) devices~\cite{emami2020ask} to help customers decide whether they should install the software applications or purchase the IoT devices.

Privacy nutrition labels aim to improve customers' awareness and help them easily learn and understand privacy practices from software applications. 
This concept was originally inspired by the common nutrition labels for food. Figure~\ref{fig:sfig1} is an example of the nutrition facts label published by the U.S. Food \& Drug Administration~\cite{foodlabel}, which demonstrates the existence and amount of indicated nutrition elements.
Moreover, the concept of privacy nutrition labels has been put into practice. 
Recently, the Google Play app store\footnote{\url{https://techcrunch.com/2022/04/26/google-play-launches-its-own-privacy-nutrition-labels-following-similar-effort-by-apple/}} and Apple App Store\footnote{\url{https://techcrunch.com/2020/12/14/apple-launches-its-new-app-privacy-labels-across-all-its-app-stores/}} also required all applications to provide their privacy nutrition labels to empower customers to learn about privacy policies before using the applications (see Figure~\ref{fig:sfig2}). 

In recent years, many researchers have extensively investigated the problem of privacy nutrition labels to assist customers in lowering their anxiety when installing software applications~\cite{ciocchetti2008future, kelley2009nutrition, li2022understanding, cranor2012necessary, balebako2014your}. Specifically, they have explored the variants of privacy nutrition labels in multiple dimensions to improve their visual information. For example, Cranor~\cite{cranor2012necessary} pointed out that privacy nutrition labels require some features, such as a standardization format, machine-readable automation, or a simple layer.
Balebako et al.~\cite{balebako2014your} showed that there is a gap between customers' interpretation and expert opinions in understanding privacy policies, leading developers to evaluate the usability of their software applications before deploying them. 
Even though privacy nutrition labels have been proven useful in the last decade, unlike food nutrition labels, there is no widely recognised and officially promulgated standard for privacy nutrition labels.
Specifically, various software application platforms have different taxonomies for the personal information involved during usage. 
Some application stores (e.g., Google Play app store) also require to include general privacy practices, such as the security measures during data transfer or whether the software application provides certain data rights (e.g., the Right to be Forgotten), in the privacy nutrition labels.

In this paper, we propose a framework for automatically generating privacy nutrition labels from verbose privacy policies to assist software application users in comprehending the privacy policies.
Specifically, by subtly utilizing the pre-trained Large Language Models (LLMs), our framework shows decent adaptability to cope with different privacy nutrition label formats.
Our framework takes as input software applications' privacy policies and platform-specific privacy label formats to construct privacy nutrition labels that satisfy the platform standards. Specifically, we first put privacy policies into the \textit{document processing module} and the \textit{context classification module} to produce privacy policy segments related to high-level data practices. These segments and platform-specific privacy label formats are then further placed into the \textit{label generation module}, which is mainly driven by an LLM, to finally generate privacy nutrition labels.
Based on our ground truth dataset, the framework achieves a 0.75 F1-score on first-party collection data practices, a 0.63 F1-score on third-party sharing data practices, and an average 0.93 F1-score on general security practices for Google's Data Safety Report format.
We summarise and attribute the wrong cases to three main reasons: omnibus data types, ambiguity caused by group-specific clauses, and under-performed context classification.
We further examine the inconsistencies between ground truth and curated privacy nutrition labels. The results show that our framework performs better for apps without privacy nutrition labels.
Our examination exposes that there could be around 12.6\% of data practices of the first-party collection are under-claimed. Our proposed framework can detect 90.1\% of these under-claimed issues.
As for Apple's App Privacy Details format, our framework yields an average 0.70 F1-score for three data practice attributes, exhibiting decent generalizability.
Moreover, we explore an alternative framework design that recursively employs another LLM to replace the first two modules that are based on Natural Language Processing (NLP) techniques. We find that the new strategy improves 13.1\% F1-score on generating third-party collection data practices, with approximately five times greater monetary cost.
Overall, our framework demonstrates decent performance in generating privacy nutrition labels from privacy policies, which can assist customers in quickly and easily understanding the privacy information contained in privacy policies.

\begin{figure*}[t!]
  \centering
  \includegraphics[width=.8\linewidth]{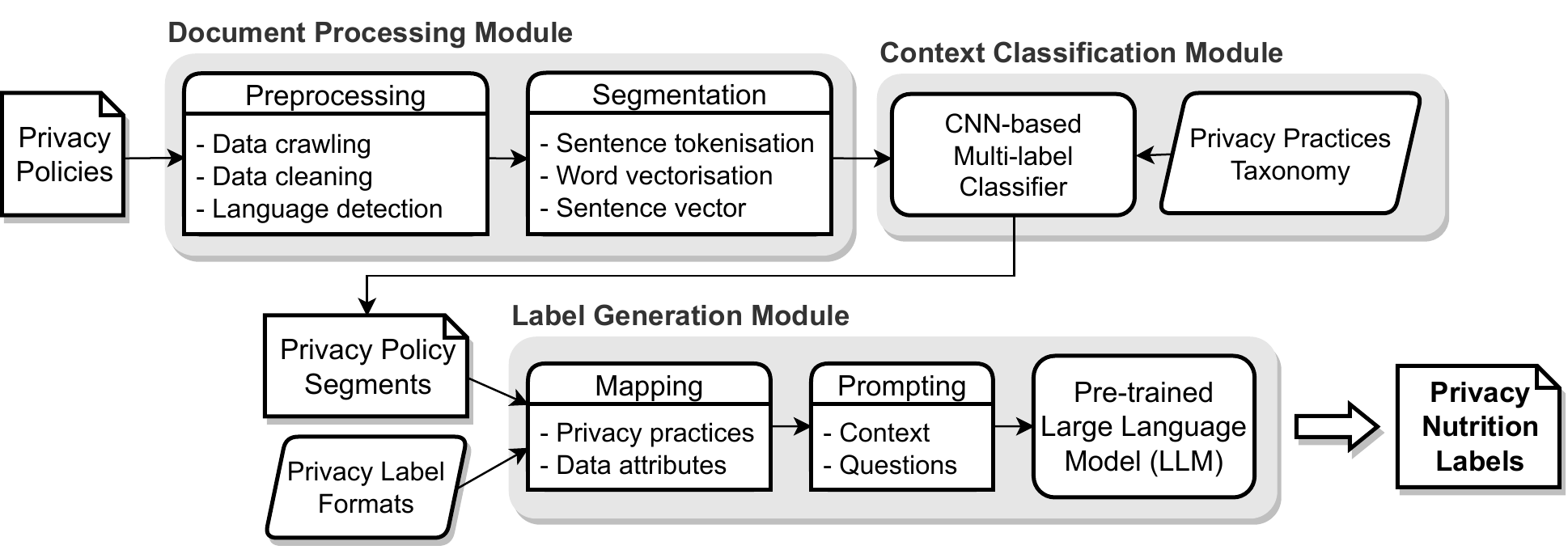}
  \caption{An overview framework for generating privacy nutrition labels from privacy policies.
  }
  \label{fig_pipeline}
\end{figure*}

The main contributions of our work are as follows: 
\begin{itemize} [leftmargin=*]
    \item To the best of our knowledge, we are the first to propose a novel framework for automatically generating privacy nutrition labels from privacy policies. 
    \item Our framework achieves respectable performance on both general and specific data practices for applications on the Google Play app store. Moreover, we also find that there are inconsistencies between the curated and ground truth privacy nutrition labels in those applications. 
    \item We employ our framework on Apple's App Privacy Details format to demonstrate its generability. The results show that our approach is applicable on multiple platforms, such as the Google Play app store and the Apple App Store. 
    \item We also propose a fully Large Language Model-based framework for constructing privacy nutrition labels from privacy policies. The results show that this framework improves the performance of data practices involving data shared with third-party. However, this framework requires much more monetary investment. 
\end{itemize}

%\shidong{Should we write like this as a separated paragraph? Or mention the Sections in-line?}
The rest of this paper is organized as follows: Section \ref{sec_methodology} introduces the motivation and structure of our proposed framework; Section \ref{sec_setup} covers the details of experimental setups, including our dataset.
Section \ref{sec_discussion} describes the results and findings of our research questions; 
Section \ref{sec_threats} discusses internal and external threats to the validity of our project; Section \ref{sec_related_work} summaries the development of related work; and lastly, in Section~\ref{sec_conclusion}, we present a conclusion and an outlook to the future.

\section{Framework of Privacy Nutrition Labels Generations}
\label{sec_methodology}

In this section, we present the motivation behind our framework design. We provide an overview of the proposed framework, which automatically generates privacy nutrition labels from privacy policies. Additionally, we introduce the three main modules of the framework in greater detail.

\subsection{Motivation}
\label{sec:motivation}

Our framework takes privacy policies as input and employs NLP techniques and pre-trained LLM, to automatically generate privacy nutrition labels for privacy policy readers. There are four main reasons to motivate the practicality of the proposed framework. 

\begin{itemize} [leftmargin=*]
    \item Nowadays, we have many existing tools that assist developers in automatically generating their privacy policies across various domains, such as websites~\cite{liu2014step, websitepolicies}, mobile applications~\cite{yu2015autoppg, zimmeck2021privacyflash, Iubenda, termly}, and e-commerce~\cite{spiekermann2001privacy}. These tools also help developers easily comply with their privacy policies with data and privacy protection laws. However, there is no tool to automatically create privacy nutrition labels to help customers easily understand privacy policies. As we have more privacy policies than privacy nutrition labels and these labels are often built based on privacy policies~\cite{kelley2009nutrition, kelley2010standardizing, kelley2013privacy}, there is a need and an opportunity to take advantage of available privacy policies to automatically construct privacy nutrition labels to save time and effort for customers, relieving the pain of reading privacy policies. 
    \item The idea of privacy nutrition labels has gained traction in recent years, with several companies and organizations adopting their own versions.
    Even though privacy nutrition labels have been proven useful in the last decade, unlike food nutrition labels, there is no widely recognised and officially promulgated standard for privacy nutrition labels.
    In addition, the laws and regulations around data privacy emerge and evolve rapidly, making it intricate to create a universal privacy nutrition labels that is meaningful and relevant for all consumers for a long period of time.
    To respond to these dynamic requirements, the automatic privacy nutrition labels generation framework is expected to have sufficient adaptability. 
    
    \item Pre-trained LLMs~\cite{floridi2020gpt, min2021recent} have shown their surprisingly strong ability on almost every task of NLP task.
    In particular, the breakthrough and far-reaching of GPT-3~\cite{brown2020language}, a pre-trained LLM, have fundamentally changed the development paradigm of deep learning models~\cite{min2021recent}.
    One of the most significant improvement of pre-trained LLMs is \textit{``One ring to rule them all.''} Specifically,
    pre-trained LLMs are trained on massive amounts of text data using unsupervised learning techniques. This means that during training, the model learns to understand the patterns and structures of language, including syntax, semantics, and contextual relationships between words and phrases. 
    These learned patterns and structures are stored in the model's weights and can be directly applied to a wide range of natural language processing tasks.
    Its strong adaptability finely matches our need; therefore, we integrate LLM into our framework to automatically generate privacy nutrition labels from privacy policies.

    \item Pre-trained LLMs have been employed in practice to solve many vital problems, such as code generation~\cite{narasimhan2021cgems, paik2021improving}, bug detection~\cite{allamanis2021self}, or user interface design~\cite{ellawela2021review}. However, if a privacy policy is directly fed into these LLMs to generate its privacy nutrition labels, there will be the following issues. First, LLMs have a limited input length, which means they will cause errors when the length of a given privacy policy is greater than the limited input length. According to~\cite{krumay2020readability}, the average length for popular apps and websites is approximately 4,000 words, while the input limitation for GPT-3 is around 1,500 English words. Second, the privacy policies are often lengthy, leading to longer processing times and requiring much more computational resources without preprocessing them. Finally, based on our preliminary results, we find that LLMs tend to return answers like, yes, presence, or positive, for yes-or-no questions along with long context prompts, which will harm the quality of generated privacy nutrition labels.
\end{itemize}
Above all, we propose a framework that basically combines traditional NLP techniques and LLMs to generate privacy nutrition labels in the desired format. 

\subsection{Framework overview}
\label{sec:overview}

Figure~\ref{fig_pipeline} illustrates the overall framework for generating private nutrition labels from privacy policies. 
The proposed framework takes a privacy policy and platform-specific privacy label format as input. The privacy policy is fed into a \textit{document processing module} and a \textit{context classification module} to produce privacy policy segments related to data practices. A list of privacy policy segments and platform-specific privacy label format is then further put into a \textit{label generation module} to generate privacy nutrition labels. The three main modules of our framework are listed as follows:  

\begin{itemize} [leftmargin=*]
    \item \textit{Document processing module}: This module takes information from the privacy policy as input and outputs a list of privacy policy segments. Following a previous work~\cite{windl2022automating}, we define that each segment may contain one to four sentences. 
    \item \textit{Context classification module}: This module takes as input a list of segments and aims to produce a list of privacy policy segments related to data practices categories. We follow the taxonomy proposed in~\cite{wilson2016creation}.
    \item \textit{Label generation module}: This module takes a list of privacy policy segments, their corresponding coarse data practices, and desired privacy label formats as inputs and outputs privacy nutrition labels by mainly employing the Q\&A function in the natural language of pre-trained LLM.
    
\end{itemize}

\noindent In the following subsections, we explain the details of each module.

\subsection{Document processing module}
\label{Sec_DocumentsProcessingModule}

Researchers often collect privacy policies using URLs~\cite{toch2010generating, apolinarski2015automating, rahmouni2014semantic}. For mobile applications, we can easily obtain privacy policy URLs by looking at their homepages on app stores, such as the Google Play app store or the Apple app store. For websites, privacy policy URLs are usually located at the top or bottom of their homepages and contain some keywords, such as ``privacy policy'', ``privacy notice'', and ``legal terms''. We employ two Python libraries, i.e., \textit{Selenium}~\cite{Selenium} and \textit{BeautifulSoup}~\cite{beautifulsoup}, to capture these privacy policies in HTML DOM format by visiting these URLs.
We then further preprocess these privacy policies
and construct their segments. We briefly present these steps in the following paragraphs.

\textbf{Preprocessing.} Following the previous work~\cite{windl2022automating}, we remove some unnecessary elements, such as script tags, CSS instructions, and headers, from privacy policy HTMLs as these elements omit actual privacy policy content. We then employ a language-detection Python library~\cite{langdetect}, namely \textit{langdetect}, to further remove texts in non-primary language from privacy policies, because some websites contain the same privacy policies in different languages. In the end, we only keep privacy policies written in English in this study, but our approach is not limited to English.

\textbf{Segmentation.} As developers often write privacy policies in lengthy and detailed texts to explain each requirement in data practices~\cite{krumay2020readability, Blakkarly2022privacy}, there is a need to group similar sentences of privacy policies into a segment to describe each specific requirement in data practices. Inspired by the previous work on the segmentation method in privacy policies~\cite{windl2022automating}, we first employ a Python natural language processing library~\cite{qi2020stanza}, namely \textit{Stanza}, to tokenize a privacy policy into multiple sentences. Second, we use a Python word embedding library~\cite{bojanowski2016enriching}, namely \textit{fastText}, to train the word embedding vector from the MAPS privacy policies datasets~\cite{zimmeck2019maps}. After this step, each token in privacy policies is represented by a word embedding vector. We then aggregate the word embedding vectors of each token in a sentence to build the sentence embedding vector. Finally, we calculate the cosine similarity between the sentence embedding vectors and merge these sentences into a segment by using a model collected from~\cite{windl2022automating}.

\textbf{At the end of this module, we obtain a list of privacy policy segments.}

\subsection{Context classification module}
\label{sec_ContextClassificationModule}

The objective of this module is to produce privacy policy segments that are highly related to data practices. We follow the previous work~\cite{wilson2016creation} to define 12 high-level categories for data practices. Specifically, the high-level categories include ``\textit{First-Party Collection/Use}'', ``\textit{Third-Party Sharing/Collection}'', ``\textit{User Access, Edit and Deletion}'', ``\textit{Data Retention}'', ``\textit{Data Security}'', ``\textit{International \& Specific Audiences}'', ``\textit{Do Not Track}'', ``\textit{Policy Change}'', ``\textit{User Choice/Control}'', ``\textit{Introductory/Generic}'', ``\textit{Practice not covered}'', and ``\textit{Privacy contact information}''.

We reuse the trained multi-label classification model mentioned in~\cite{windl2022automating}. Specifically, the multi-label classification model contains a convolutional neural network, including one convolutional layer, one pooling layer, and two dense layers. 
Based on previous studies~\cite{harkous2018polisis, windl2022automating}, we set the probability threshold of our classification model at 0.5. Typically, if the predicted probability score of a privacy segment is greater than 0.5, the privacy segment is labeled as the corresponding data practice(s). 

\textbf{At the end of this module, we can obtain privacy policy segments that are highly related to the categories of data practices.}

\subsection{Label generation module}
\label{sec_LabelsGenerationModule}

The target of this module is to accurately and efficiently generate privacy nutrition labels. The module takes privacy policy segments, their corresponding data practices categories, and privacy label format as inputs and outputs privacy nutrition labels by mainly employing the LLM, i.e., GPT-3. We describe how we process inputs step-by-step in the following paragraphs.

\textbf{Privacy nutrition labels format.} To standardise the displayed information on privacy nutrition labels, platforms usually have a set of requirements that specifically define the information the developers need to provide.
Similar to the food nutrition label, which has different sections such as trace elements, vitamins, and energy, privacy nutrition labels also explicitly categorise essential information into various sections.
And for each section, a set of attributes needs to be fulfilled by the developers.
For example, as we shown in Table~\ref{tab_google_label}, Google stipulates three sections for the Data Safety Report: \textit{first-party data collected}, \textit{data shared with third-party}, and general \textit{security practices}; and for each section, attributes, and their scopes are also designated.

\textbf{Mapping.} After obtaining the privacy nutrition labels format, we map each section with one or more high-level data practices categories discussed in the \textit{context classification module}. For example, [``\textit{first-party collection/use}'' and \textit{first-party data collected}]; [``\textit{user access, edit and deletion}'', ``\textit{data retention}'' and \textit{RTBF}]. By doing so, only related privacy policy segments will be fed into the LLM.

\textbf{LLM and prompt.} The pre-trained LLM requires users to provide a \textit{prompt} to identify the context of a given question and the question itself for Q\&A completion. According to Floridi and Chiriatti~\cite{floridi2020gpt}, the results of LLMs highly depend on the definition of the prompt, i.e., clear, reasonable, and appropriate. Moreover, some studies show that we could amplify LLMs' problem-solving capabilities by providing good prompts~\cite{kojima2022large, gpt3close}. For this reason, we aim to design our prompts based on expected privacy nutrition labels format.
As for the question design, LLMs show stronger capability on open-ended question answering compared to closed-ended questions~\cite{hendrycks2020measuring, gpt3close}.
However, the intuitive open-ended Q\&A inevitably requires more effort in processing and analysing diverse answers.
On privacy nutrition labels, to guarantee the uniformity of displayed information, all attributes' values are from a predefined universal set, therefore we believe that close-ended Q\&A fits our task better.
To alleviate the inherent weaknesses of LLM in answering closed-ended questions, we design our questions in a yes-or-no format rather than closed-ended multi-choice questions. 
As for the context, we directly use the related privacy policy segments based on the mapping. Moreover, we also include the application's name displayed in the market in the context to improve the LLMs' performance.
By combining the question and its context as prompt, we then iterate the question-answer process on all privacy nutrition label attributes and collect answers. 
If the answer starts with \textit{``yes''}, we count the value for this attribute as \textit{presence} or \textit{positive}, otherwise \textit{absence} or \textit{negative}. 
Detailed explanations of prompt design are presented in the supplementary material. 

\textbf{At the end of this module, we can obtain the generated privacy nutrition labels.}

\section{Experimental Setup}
\label{sec_setup}

In this section, we describe our dataset and evaluation metrics used to estimate the performance of our framework. We also justify the selection of the pre-trained LLM.

\begin{table}[t!]
\centering
\caption{\faIcon{google} Data safety reports format on the Google Play app store. \textit{RTBF} stands for the Right To Be Forgotten.}
\label{tab_google_label}
\resizebox{1.0\linewidth}{!}{%
\begin{tabular}{llr} 
\toprule
\multicolumn{3}{c}{\textbf{First-party data collected}}                                         \\ 
\midrule
\textbf{Attribute} & \textbf{Description}                          & \textbf{Attribute value}   \\ 
\midrule
Data type          & The data type that the app claims to collect. & 38 types                   \\ 
\midrule
\midrule
\multicolumn{3}{c}{\textbf{Data shared with third-party} }                                  \\ 
\midrule
\textbf{Attribute} & \textbf{Description}                          & \textbf{Attribute values}  \\ 
\midrule
Data type          & The data type that the app claims to share.   & 38 types                   \\ 
\midrule
\midrule
\multicolumn{3}{c}{\textbf{Security practices}}                                                 \\ 
\midrule
\textbf{Attribute} & \textbf{Description}                          & \textbf{Attribute values}  \\ 
\midrule
Encryption         & Data is encrypted in transit.                  & Yes/No                     \\ 
\midrule
\textit{RTBF}              & You can request that data be deleted.          & Yes/No                     \\
\bottomrule
\end{tabular}
}%
\vspace{-5pt}
\end{table}

\subsection{Dataset}
\label{sec_dataset}

To evaluate the performance of our framework, we need to construct the ground truth of privacy nutrition labels. The Google Play app store is one of the pioneering stores requiring its applications to provide privacy nutrition labels~\cite{li2022understandingios}. Thus, we collect Android mobile applications with their privacy policies and privacy nutrition labels from the Google Play app store. In the following paragraphs, we present the details of its privacy nutrition labels format and our collection strategy.

\textbf{Google Play app store and its Data Safety Report.} 
Android mobile applications are widely discussed and recognized as transparent and friendly to academic research~\cite{li2022understandingios, USENIX_2022_GEODIFF}. Although there are many application markets for the Android platform, the Google Play app store is the largest (with over 2.6 million applications) and most accessible app market~\cite{appbrain}. Moreover, the Google Play app store not only requires developers to provide a privacy policy link for their applications but also the Data Safety Report, which is a form of privacy nutrition labels.
We summarize the Data Safety Report format (privacy nutrition labels) in Table~\ref{tab_google_label}. 
The Data Safety Report~\cite{googledatasafety} contains privacy information from three perspectives. The first section is the \textit{first-party data collected}, which includes 38 data types. Similarly for the section for \textit{data shared with third-party}, which also includes 38 data types. For \textit{security practices}, the \textit{encryption} and the \textit{Right To Be Forgotten} (RTBF) attributes give information about whether data is encrypted during transit and whether the users can request their data to be deleted. 

\begin{figure}[t!]
 \includegraphics[width=0.8\linewidth]{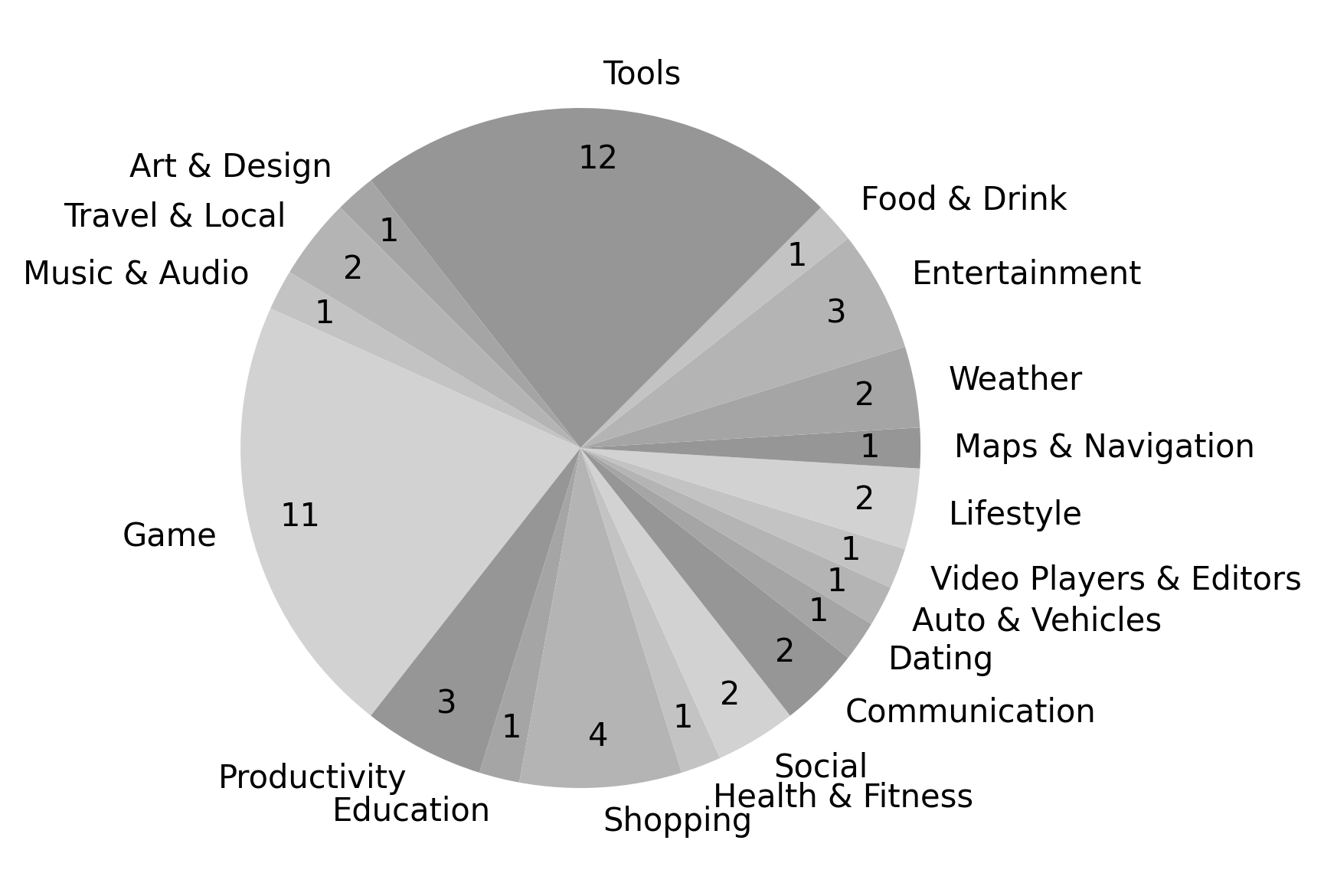}
 \caption{The application category distribution of our dataset.}
 \label{fig_app_distribution}
\vspace{-5pt}
\end{figure}

\textbf{Apps selection.} We randomly harvest around 10\% of mobile applications from the Google Play app store based on the AndroZoo collection~\cite{androzoo}. We found that only 17,675 \textbf{\ul{(6.6\%)}} mobile applications provide a Data Safety Report, whilst 84.3\% of applications provide a privacy policy on their Google Play app store homepages. We suspect that the Google Play app store has not rigorously enforced the new requirements for privacy nutrition labels, which further raises our concerns about the quality of existing privacy nutrition labels. According to previous studies~\cite{CMU_2017, TSE_2018, zimmeck2019maps, liu2021have}, we often have a large number of mobile applications containing low-quality privacy policies; hence, we should remove these privacy policies so we can accurately generate privacy nutrition labels. To collect high-quality privacy policies from mobile applications, we follow previous work~\cite{zimmeck2019maps, liu2021have, USENIX_2022_GEODIFF}. Specifically, we first remove privacy policies that less than 200 words~\cite{USENIX_2022_GEODIFF} or file size smaller than 2KB~\cite{liu2021have}. Then we refer to the rate of applications, the number of installations, popular categories, etc., to select mobile applications having high-quality privacy policies. In the end, our dataset collection contains 52 mobile applications. Figure~\ref{fig_app_distribution} describes the application category distribution of our dataset. We then further extract their privacy policies and manually label their privacy nutrition labels.

\begin{table}[t!]
  \caption{The statistical analysis of our dataset.}
  \label{tab_dataset_summary}
  \begin{tabular}{l | c}
    No. Privacy policies & 52  \\
    No. Segments & 7,659 \\
    No. Words & 202,932\\
    \hline
    \hline
    Segments per Privacy policy & 151 \\
    Words per Segment & 27\\
    Words per Privacy policy & 3,979\\
\end{tabular}
\end{table}

\textbf{Ground truth dataset.} 
There has been no academic research done to ensure the quality of privacy nutrition labels on the Google Play app store.
Additionally, according to a recent industrial report from Mozilla~\cite{SeeNoEvil}, nearly 80\% of the apps they reviewed have some discrepancies between the apps’ privacy policies and the information they claimed on Google’s Data Safety Report.
Moreover, only a small portion of applications have privacy nutrition labels, leading to deep concern about their quality. For these reasons, we need to create a ground truth dataset for privacy nutrition labels as in the data safety report format to accurately evaluate our framework. We recruited two Ph.D. students with privacy policy research backgrounds to read the original privacy policies and label all attributes in each section individually. For any disagreement, the students discussed and agreed on the same answer, and if the disagreement persisted, a third author (a senior researcher) joined the discussion to facilitate a resolution. Table~\ref{tab_dataset_summary} further presents the statistical analysis of our collected privacy policies in terms of the number of segments and the number of words.

\subsection{Evaluation metrics}
\label{sec:metrics}

We employ classification evaluation metrics such as precision, recall, and F1-score to estimate the performance of our framework. Specifically, we aim to guarantee the uniformity of privacy nutrition labels, i.e., all attributes come from a universal attribute set and all attributes' values are also from predefined universal attribute sets (see Table~\ref{tab_google_label}). We use the macro-average to calculate the scores for each evaluation metric to prevent a bias in our results.

\subsection{Large language model selection}
Language models have seen tremendous advancements in recent years, especially with numerous large models such as GPT-3~\cite{brown2020language}, chatGPT~\cite{chatGPT}, GPT-4~\cite{openai2023gpt4}, LLaMA~\cite{touvron2023llama}, and Bard~\cite{GoogleBard}.
We select GPT-3 in this paper for the three following reasons: First, GPT-3 is one of the earliest LLMs published in 2020, and basically, all following LLMs are proved stronger than GPT-3 on all dimensions. Thus, we believe various LLM selections will not negatively affect the fidelity and performance of our framework. Second, GPT-3 has been the focus of extensive research and analysis by the academic community for a significant period. This indicates that there is a wealth of knowledge and resources available for further analysis and investigation in the future. Third, some researchers have proposed and proved that GPT-3 could be compressed into a much smaller model while maintaining similar performance~\cite{shi2022compressing, tao2022compression}.
The slimming of LLMs enables the possibility that our framework is extended to work offline for some specific IoT scenarios.

\section{Results and Discussion}
\label{sec_discussion}

\begin{table*}[t!]
  \caption{\faIcon{google} Performance of our proposed framework on the ground truth dataset (RQ1). \textit{RTBF} stands for the Right To Be Forgotten.}
  \label{tab_google_results_1}
  \vspace{-5pt}
  \subfloat[Ground truth dataset \textit{with} all data types 
  % \shidong{should we add the $\mu$ to represent the average for all apps?}
  ]{
  \begin{tabular}{l r r r}
    \toprule
     & \textbf{Prec.} & \textbf{Rec.} & \textbf{F1-score} \\
        \midrule
    \textbf{First-party data collected} & 0.772 & 0.797 & 0.751\\
        \midrule
    \textbf{Data shared with third-party} & 0.668 & 0.753 & 0.632\\
        \midrule
    Encryption & /  & / & 0.942\\
    \textit{RTBF} & / & / & 0.923 \\
    \midrule
    \textbf{Security practices} & / &/ & 0.932 \\
    
  \bottomrule
\end{tabular}
}%
 \quad 
  \subfloat[Ground truth dataset \textit{without} 14 omnibus data types]{
  \begin{tabular}{l r r r}
    \toprule
    & \textbf{Prec.} & \textbf{Rec.} & \textbf{F1-score} \\
    \midrule
    \textbf{First-party data collected}  & 0.773 & 0.814 & 0.758\\
        \midrule
    \textbf{Data shared with third-party}  & 0.689 & 0.781 & 0.655\\
        \midrule
    Encryption & /  & / & 0.942\\
    \textit{RTBF} & / & / & 0.923 \\
    \midrule
    \textbf{Security practices} & / &/ & 0.932 \\
  \bottomrule
\end{tabular}
}%
\vspace{-5pt}
\end{table*}

In this section, we present and discuss the results of the following research questions (RQs):

\begin{itemize} [leftmargin=*]
    \item \textbf{RQ1:} What is the performance of the proposed framework?
    \item \textbf{RQ2:} Are there any inconsistencies between the ground truth and the privacy nutrition labels curated on the Google Play app store?
    \item \textbf{RQ3:} How is the generalizability of the proposed framework?
    \item \textbf{RQ4:} What is the performance of fully LLM-based design for the proposed framework? 
\end{itemize}

\subsection{Performance of our framework (RQ1)}
\label{sec_RQ1}

Our results from the ground truth dataset are shown in Table~\ref{tab_google_results_1}a. The results show the overall \textit{first-party data collected} is better than the \textit{data shared with third-party}. We find that privacy policies tend to clearly enumerate the data practices for first-party, but indefinitely for third-party. 
This observation is intuitively reasonable and conforms to conclusions from previous studies~\cite{SeeNoEvil, li2022understanding, li2022understandingios}.
Furthermore, both attributes in \textit{security practice} achieve over 0.9 F1-score (accuracy) , demonstrating strong capability in the general data practices of our framework. We further manually inspect and analyze the wrong cases, summarising the following reasons behind them.

\textbf{Omnibus data types.} In law, clauses that include the term ``other(s)'' are often referred to as ``omnibus'' or ``catch-all'' clauses.
As for the Data Safety Report format presented by the Google Play app store, we notice that there are seven data types for \textit{first-party data collected} and seven data types for \textit{data shared with third-party} include the term ``other'' in their names or definitions.
For example, \textit{Other Info} which is defined as ``\textit{Any other personal information, such as date of birth, gender identity, veteran status, etc.}'', or
\textit{Other In-app Messages} which is represented as ``\textit{Any other types of messages. For example, instant messages or chat content.}''
The use of the omnibus data types can be helpful in situations where the privacy nutrition labels format is intended to be flexible or adaptable, as it allows for the inclusion of new or unforeseen circumstances that may arise over time. 
However, these data types lead to ambiguity or uncertainty about their scope and application to the customers. They also raise the understanding difficulty of our framework. 
According to Table~\ref{tab_google_results_1}b, the performance of our framework increases on all evaluation metrics, i.e., precision, recall, and F1-score, if we exclude the 14 omnibus data types.

\textbf{Ambiguity caused by group-specific clauses.}
To be compatible and compliant against regulation laws such as GDPR~\cite{GDPR} and CCPA~\cite{CCPA}, privacy policies should have some group-specific clauses that are only applicable to certain groups of users based on residency or age. Therefore, privacy policies commonly have group-specific clauses to respond to various regulatory requirements, which brings ambiguity issues to our generation process.
Since the target user group is not specified, our model will be confused by some contradictory group-specific clauses on the same privacy or data practice.
For example, many privacy policies mentioned that they do not collect or share any data from underage users, while they still actively collect or share data from adult users.
In this case, it does not matter whether our model returns presence or absence for a data practice; either should be counted as correct. 
Therefore, group-specific privacy nutrition labels are expected to be explicitly defined by the platform, and with the clear group-specific report format definitions and requirements as the input, our model would exhibit better performance.

\textbf{Under-performed context classification.}
Another significant factor that limits our model's performance is the capability of the \textit{context classification module}. The average F1-score of this module used to predict the high-level categories of data practices is 0.70. The worst class among the 12 high-level data practices is \textit{``Data Retention''}, with a 0.31 F1-score.
For example, the app Super Slime Simulator~\cite{SuperSlimeSimulator}, a creative game app, we find the segment including the following sentence, i.e., \textit{``You can request the deletion of the Personal Information. ...withdraw your consent to the processing of such information, and disable its future collection by contacting us at...''}, which is incorrectly classified as \textit{``Third-party Sharing/Collection''}, instead of \textit{``Data Retention''}.
Consequently, this segment is not mapped with the attribute \textit{RTBF} in the security practices section, and is not fed into the \textit{label generation module}.
As a result, the \textit{RTBF} of this app is wrongly labeled as ``absence.'' We manually put this segment into the \textit{label generation module}, and our module labels the segment as \textit{RTBF}.
Therefore, it is essential to improve the classifier performance in the \textit{context classification module} to further enhance the performance of our proposed framework.

\begin{tcolorbox}
  \textbf{Finding 1:} 
  Our framework demonstrates decent performance on general data practices, yielding 0.94 and 0.92 in terms of F1-score for encryption and \textit{RTBF}, respectively. As for specific data practices, such as first-party data collected and data shared with third-party, our framework achieves 0.75 and 0.63 F1-scores, respectively. 
\end{tcolorbox}

\subsection{Discrepancy between ground truth and curation on market (RQ2)}
\label{sec_RQ2}
\begin{table*}[htbp]
  \caption{\faIcon{google} Performance of our proposed framework for apps without a data safety report on the Google Play app store (RQ2). \textit{RTBF} stands for the Right To Be Forgotten.}
  \label{tab_google_results_2}
  \vspace{-5pt}
  \subfloat[Ground truth dataset \textit{with} all data types]{
  \begin{tabular}{l r r r}
    \toprule
    & \textbf{Prec.} & \textbf{Rec.} & \textbf{F1-score} \\
    \midrule
    
    % 0.7934987612757165, 0.8657581505162923, 0.8079313487026298
    \textbf{First-party data collected} & 0.794 & 0.866 & 0.808\\
    \midrule
    % 0.6876822714322715, 0.7863399956866384, 0.6747971282019489
    \textbf{Data shared with third-party} & 0.688 & 0.786 & 0.675\\

    \midrule
    Encryption & /  & / & 1.000 \\
    \textit{RTBF} & / & / & 1.000 \\
    \midrule
    \textbf{Security practices} & / &/ & 1.000 \\
  \bottomrule
\end{tabular}
}%
 \quad 
  \subfloat[Ground truth dataset \textit{without} 14 omnibus data types]{
  \begin{tabular}{l r r r}
    \toprule
    & \textbf{Prec.} & \textbf{Rec.} & \textbf{F1-score} \\
    \midrule
    % 0.7977288427946323, 0.8848898685855209, 0.8106283167868534
    \textbf{First-party data collected}  & 0.798 & 0.885 & 0.811\\
    \midrule
    % 0.6876933073424302, 0.8721983420732031, 0.6829566965205663
    \textbf{Data shared with third-party}  & 0.688 & 0.872 & 0.683\\
    \midrule
    Encryption & /  & / & 1.000 \\
    \textit{RTBF} & / & / & 1.000 \\
    \midrule
    \textbf{Security practices} & / &/ & 1.000 \\
  \bottomrule
\end{tabular}
}%
\end{table*}

In Section~\ref{sec_dataset}, we find that a substantial proportion of applications fail to provide privacy nutrition labels on the Google Play app store. Thus, we take a closer look at our dataset and notice that nine applications do not claim any data practices, including five tool applications, two game applications, one productivity application, and one communication application. However, based on the ground truth dataset, these applications mention at least one data practice in their privacy policies. For example, Easy Booster\footnote{\url{https://play.google.com/store/apps/details?id=easy.booster.clean.tool}} is a tool application to make mobile phones run smoother and faster, with over 10 million installs. Its privacy policy contains the following phrase: \textit{``Information Collection and Use ... Log Data. We want to inform you that whenever you use our Service ... This Log Data may include information such as your device Internet Protocol (“IP”) address, device name, operating system version, the configuration of the app when utilizing our Service ...''}

Table~\ref{tab_google_results_2}a and Table~\ref{tab_google_results_2}b show the performance of the applications with and without the 14 omnibus data types, respectively.
The results indicate a marginal improvement over the mean performance of the entire dataset, namely 0.05 (+7.9\%) F1-score for \textit{first-party data collected}, 0.04 (+6.8\%) F1-score for \textit{data shared with third-party}, and 0.07 (+7.2\%) F1-score for \textit{security practices}.
Various privacy policy generation tools are available for software applications, which include source code analysis-based tools~\cite{yu2015autoppg, TSE_2018, zimmeck2021privacyflash} and online automated privacy policy generators that employ questionnaires~\cite{Iubenda, termly, websitepolicies}.
However, there are no existing tools that can assist developers in constructing privacy nutrition labels from privacy policies.
Our promising results suggest that our framework can be applied to further assist them in automatically generating privacy nutrition labels, especially for the data practices related to first-party or shorter privacy policies.

\begin{figure}[t!]
 \includegraphics[width=1\linewidth]{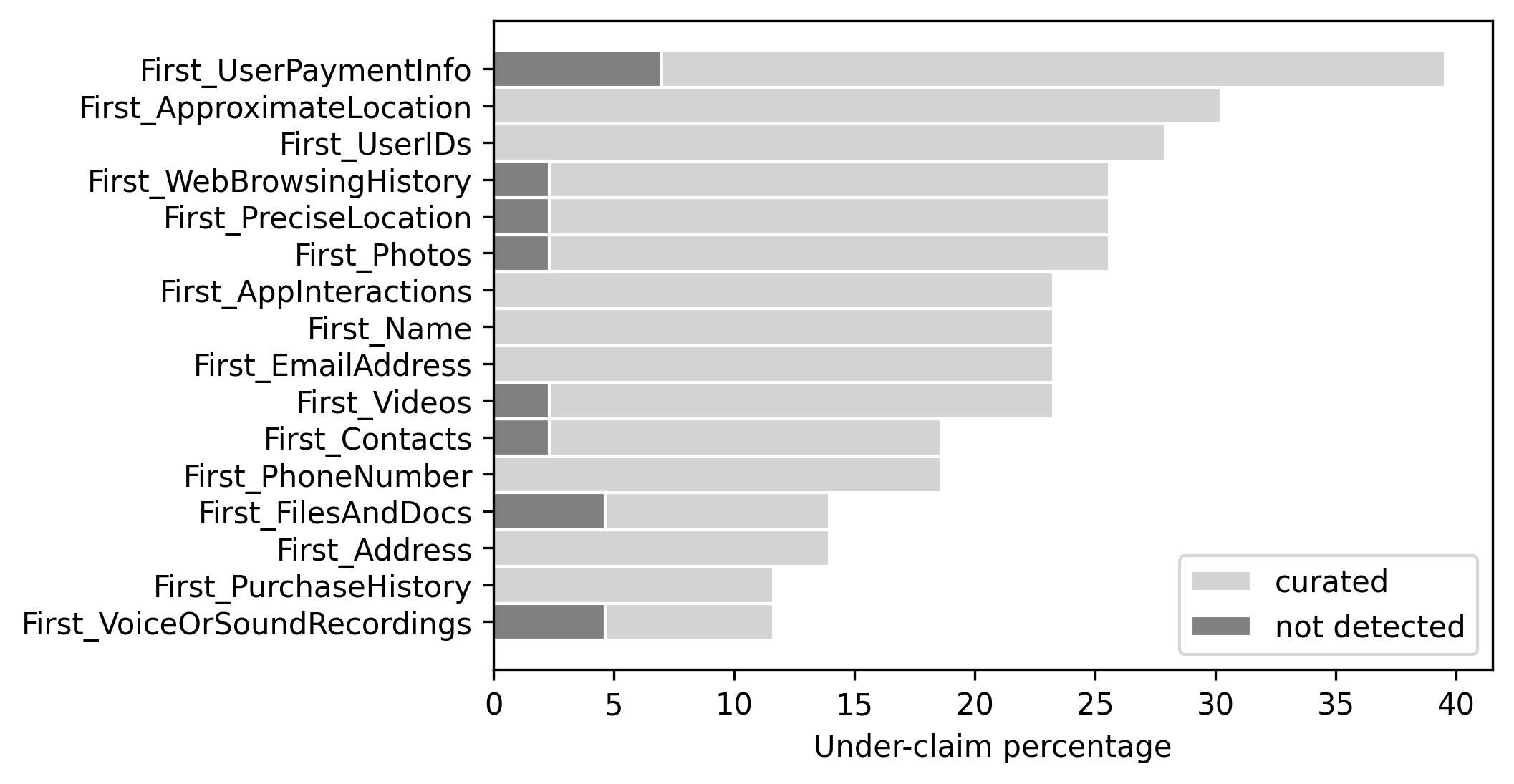}
 \vspace{-10pt}
 \caption{For privacy nutrition labels curated on the market, the under-claim rate of \textit{first-party data collected} in percentage (as light grey bars) and the not detected under-claim rate (as dark grey bars). Inclusive attributes are removed. Attributes that are less than 10\% under-claim rate are omitted.}
 \label{fig_underclaim}
 \vspace{-10pt}
\end{figure}

\textbf{Under-claimed data practices.}
Several previous studies have discussed and identified common challenges developers when they construct privacy nutrition label~\cite{li2022understandingios, li2022understanding}. Our findings are similar to theirs. 
Specifically, except for the absence of privacy nutrition labels, we find that there are many under-claimed data practices in the Data Safety Report curated on the Google Play app store, compared to their privacy policies.
For example, Match Masters~\cite{MatchMasters} is a tile-matching game app with over 10 million installs. 
In their privacy policy, they wrote a long list of \textit{``Which Information do we collect?''}, including \textit{``Personal Data (optional)'', ``Location'', ``Sessions log'', and ``Mobile device model''}. We are able to identify all corresponding first-party data practices in its Data Safety Report \textbf{except} \textit{Location}.
Although there are four conditions under which developers are exempt from disclosing data accessed by an application as ``collected'' in the Data Safety Report, there is no special explanation about \textit{``Location''} in the privacy policy. Thus, we regard the missing \textit{``Location''} as an under-claim. 
In total, there are around 12.6\% of data practices, related to first-party data collected, are under-claimed, excluding omnibus data types. The specific under-claim rates for each data type are shown in Figure~\ref{fig_underclaim} as light gray bars. The top three under-claimed data types are \textit{User Payment Info}, \textit{Approximate Location}, and \textit{User IDs}.

\textbf{Under-claiming detection.}
Under-claiming data practices in privacy information undermines user trust, can lead to unintended disclosure of privacy information, and may have legal consequences. To tackle this challenging issue, we employ our framework for detecting under-claim issues. The results show that our framework is able to identify 90.1\% of these issues. Especially, for data practices about basic personal information, such as \textit{User IDs}, \textit{Name}, \textit{Email Address}, \textit{Phone Number}, and \textit{Address}, our framework can successfully notice 100\% of under-claim issues (See Figure~\ref{fig_underclaim}).
Generally speaking, individually verifying all under-claiming is too complicated to conduct without specific legal knowledge, as many cases merely fit the exemption conditions. Still, in addition to providing early reminders to the customers, our framework is applicable to assist legal experts in cross-checking and detecting the consistency between privacy policies and their privacy nutrition labels.

\begin{tcolorbox}
  \textbf{Finding 2: } 
  There are inconsistencies between the generated and ground truth privacy nutrition labels. Specifically, around 17\% of applications do not provide privacy nutrition labels and for those applications, our framework can achieve 0.81, 0.68, and 1.00 in terms of F1-score for first-party data practices, third-party data practices, and security practices, respectively. For the Data Safety Report curated on the Google Play app store, there are 12.6\% under-claim data practices. Our framework can detect \textbf{90.1\%} of them.
\end{tcolorbox}

\subsection{Generalizability of proposed framework (RQ3)}
\begin{table*}[t!]
  \caption{\faIcon{apple} Performance of our proposed pipeline on Apple's app privacy details. (RQ3)}
  \label{tab_apple_result}
  \vspace{-10pt}
\centering
\subfloat[All available apps]{
  \begin{tabular}{l r r r}
    \toprule
    Attribute & \textbf{Prec.} & \textbf{Rec.} & \textbf{F1-score} \\
    \midrule
    % [0.7978021978021981, 0.7629931972789116, 0.777620474763332, 0.7373836546476297]
    % total_notLinked_results: 
    % [0.8439560439560442, 0.7688600685029255, 0.799178797393083, 0.6866976068840961]
    % total_tracked_results: 
    % [0.7318681318681323, 0.7685977514548943, 0.7010694464265893, 0.648190640235974]

    \textbf{Data linked to you}  & 0.763 & 0.778 & 0.737\\
    \textbf{Data not linked to you} & 0.769 & 0.799 & 0.687\\
    \textbf{Data used to track you} & 0.769 & 0.701 & 0.648\\
    \midrule
    \textbf{Average} & 0.767 & 0.759 & 0.691\\
  \bottomrule
\end{tabular}
}%
\quad 
  \subfloat[All available apps with curated privacy nutrition labels]{
    \begin{tabular}{l r r r}
    \toprule
    Attribute & \textbf{Prec.} & \textbf{Rec.} & \textbf{F1-score} \\
    \midrule
    % total_linked_results: 
    % [0.7980769230769235, 0.7816964285714286, 0.7712527056277056, 0.7563490347967249]
    % total_notLinked_results: 
    % [0.8509615384615388, 0.7671893644549894, 0.811080976315351, 0.6910639730113668]
    % total_tracked_results: 
    % [0.7259615384615388, 0.7649790747446997, 0.6960915820290821, 0.6372137210914299]
    \textbf{Data linked to you}  & 0.782 & 0.771 & 0.756\\
    \textbf{Data not linked to you} & 0.767 & 0.811 & 0.691\\
    \textbf{Data used to track you} & 0.765 & 0.696 & 0.637\\
    \midrule
    \textbf{Average} & 0.771 & 0.759 & 0.695\\
  \bottomrule
\end{tabular}
}%
\end{table*}

\begin{table}[t!]
\centering
\caption{\faIcon{apple} Apple's app privacy details format. Notably, the first two attributes are not mutual exclusive.} 
% \zc{put the tables/figures always at the top of the page}
\vspace{-6pt}
\label{tab_apple_label}
\resizebox{1.0\linewidth}{!}{%
\begin{tabular}{llr} 
\toprule
\multicolumn{3}{c}{\textbf{Data type}}                                         \\ 
\midrule
\textbf{Attribute} & \textbf{Description}                          & \textbf{Attribute Value}   \\ 
\midrule
Data linked to you          & \makecell[l]{Data that is linked to your\\ identity (via your account, \\ device, or other details).} & 13 categories                  \\ 
\midrule
Data not linked to you         & \makecell[l]{Data that is not linked to your \\ identity (via your account, \\ device, or other details).}  & 13 categories                  \\ 
\midrule

Data used to track you         & \makecell[l]{Data may be used to track you \\ across apps and websites \\ owned by other companies}               & 13 categories                    \\ 
\bottomrule
\end{tabular}
}%
\vspace{-10pt}
\end{table}

To demonstrate the generalizability of our method, we further employ it to generate privacy nutrition labels in Apple's App Privacy Details format. The Apple App Store is Apple's official app market for mobile apps on its iOS and iPadOS operating systems. Similar to Google's Data Safety Report, Apple's version of privacy nutrition labels is called App Privacy Details. Developers are required to provide this information to submit new applications and their updated information to the Apple App Store.
Compared to Data Safety Report in the Google Play app store, there are two prominent format differences for Apple's App Privacy Details (See Table~\ref{tab_apple_label}). First, App Privacy Details classify all data practices into three categories: \textit{data linked to you}, \textit{data not linked to you}, and \textit{data used to track you}; Second, Apple's App Privacy Details do not have general security practices that are applicable to all data practices.

Among the 52 applications in the original ground truth dataset, 13 applications are not published on the Apple App Store and four applications have various privacy policy links. We remove these applications, and there are 35 remaining applications. We employ our framework for these privacy policies' applications to generate privacy nutrition lables as per Apple's App Privacy Details format. Table~\ref{tab_apple_result}a shows the performance of our framework on the 35 applications. We also further exclude three applications that do not provide privacy nutrition labels. Table~\ref{tab_apple_result}b presents the results of our framework for the remaining 32 applications.

The information included in Apple's App Privacy Details is more challenging to comprehend than the privacy policies. Especially when links between data practices and users are unknown, the process of generating privacy nutrition labels or drafting privacy policies is more challenging~\cite{li2022understanding, li2022understandingios, zhang2022usable}. 
The results of our evaluation demonstrate that the framework has the best performance in generating privacy nutrition labels for \textit{data linked to you}, followed by \textit{data not linked to you} and \textit{data used to track you}. This ranking aligns with the findings of previous studies and corresponds to our intuitive sense of the relative difficulty of understanding these attributes.
Our framework achieves overall comparable results, i.e., 0.77 precision, 0.76 recall, and 0.70 F1-score, showing that the proposed framework performs decent generalizabiltiy on generating privacy nutrition labels in various formats.

\begin{tcolorbox}
  \textbf{Finding 3:} Our framework exhibits decent generalizability by achieving an overall 0.77 precision, 0.76 recall, and 0.70 F1-score on Apple's App Privacy Details format.
\end{tcolorbox}

\subsection{A trial of fully LLM-based solution (RQ4)}
\begin{table*}[htbp]
  \caption{\faIcon{google} Performance of an alternative framework design based on recursively employing LLMs (RQ4).}
  \label{tab_google_results_3}
  \vspace{-5pt}
  \subfloat[Ground truth dataset \textit{with} all data types]{
  \begin{tabular}{l r r r}
    \toprule
    & \textbf{Prec.} & \textbf{Rec.} & \textbf{F1-score} \\
    \midrule
     % 0.7570780005019135, 0.7869673474673474, 0.7443452168069195]
    \textbf{First-party collected} & 0.757 & 0.787 & 0.744\\
    \midrule
    % 0.7585376510462718, 0.7606073656073657, 0.714985200427342
    \textbf{Shared with third-party} & 0.759 & 0.761 & 0.715\\
  \bottomrule
\end{tabular}
}%
 \quad 
  \subfloat[Ground truth dataset \textit{without} 14 omnibus data types]{
  \begin{tabular}{l r r r}
    \toprule
    & \textbf{Prec.} & \textbf{Rec.} & \textbf{F1-score} \\
    \midrule
    %  0.7548221824615015, 0.8109307359307361, 0.7512735660076695
    \textbf{First-party collected}  & 0.755 & 0.811 & 0.751\\
    \midrule
    % 0.7763364103577055, 0.7568953089920832, 0.7069601530109569
    \textbf{Shared with third-party}  & 0.776 & 0.757 & 0.710\\
  \bottomrule
\end{tabular}
}%
\end{table*}
Chain-of-AI (CoAI) or Chain-of-Thought (CoT) has gained significant popularity in recent several years as a means of creating more powerful and sophisticated AI systems.
By combining multiple AI models, especially LLMs, in a sequential or interconnected manner, this approach has been shown to be effective in natural language processing tasks, such as machine translation and chatbots~\cite{gulcehre2017integrating, schwenk2008large}, amplifying LLMs through recursive question-answering and debate~\cite{mialon2023augmented}.
However, some arguments~\cite{why2023yang} suggest that for relatively simpler natural language understanding (NLU) tasks, LLMs may not outperform traditional fine-tuned models.
As we discussed in Section~\ref{sec_RQ1}, under-performed \textit{context classification module} limits the performance of framework. Inspired by those ideas, we propose a fully LLM-based framework design to explore its performance.

\begin{figure}[t!]
 \includegraphics[width=1\linewidth]{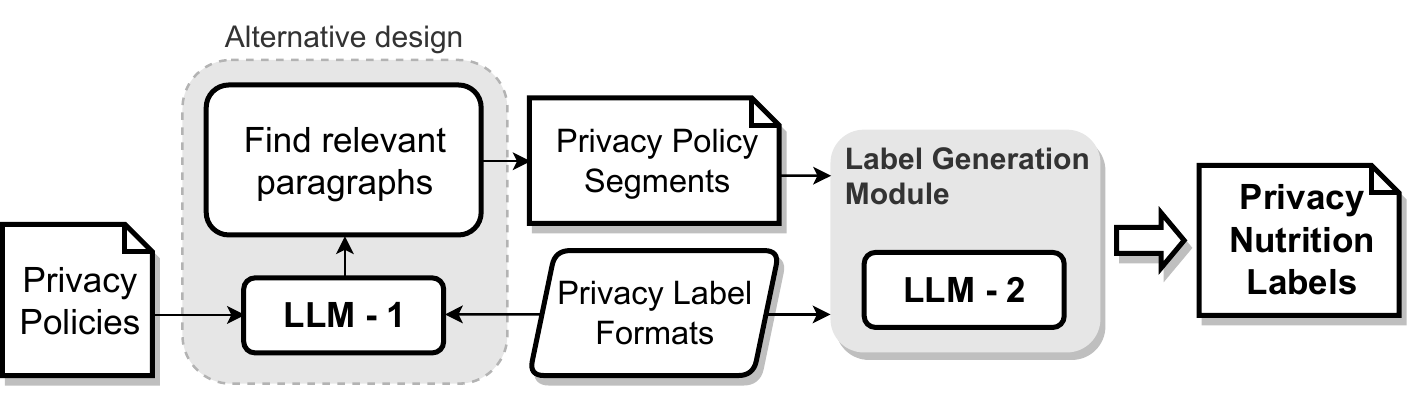}
 \caption{An overview for the fully LLM-based framework.}
 \label{fig_alternative}
\end{figure}

Figure~\ref{fig_alternative} shows an overview of our alternative framework. As we discussed in Section~\ref{sec_methodology}, the \textit{document processing module} and the \textit{context classification module} are implemented based on traditional NLP techniques, and the output is privacy policy segments with their privacy practice labels.
For alternative design, we employ another LLM (LLM-1) to conduct the equivalent task.
Specifically, for every question that will be processed by the LLM-2 in the \textit{label generation module}, we employ LLM-1 to find its relevant paragraphs from the privacy policy documents. Detailed explanations and prompt examples are presented in the supplementary material.

Table~\ref{tab_google_results_3} shows the performance of our alternative framework design. Notably, given the approximately same LLM usage quota, we were only able to run around 20\% of the dataset by random sampling. In addition, since \textit{security practices} have obtained an average over 0.9 F1-score, we skip them in this research question. This new framework design achieves approximate performance on all metrics for \textit{first-part data collected} but improves 0.09 (+13.6\%) and 0.08 (+13.1\%) on precision and F1-score for \textit{data shared with third-party}, respectively. However, there is no noticeable increase for data practices involving the \textit{first-party data collected}.
Compared to the original \textit{document processing module} and the \textit{context classification module} based on traditional NLP techniques, the extra LLM can better collect and perceive third-party information that is usually scattered across the privacy policy document.

On the other hand, the cost of implementing this approach is a notable side effect, since most mainstream LLMs are pay-to-use based on input length and we recursively employ them to process long privacy policy documents.
Based on our observation, we coarsely estimate the monetary cost of fully LLM-based solution at more than five times that of the original framework design.

\begin{tcolorbox}
  \textbf{Finding 4:} 
  A fully LLM-based framework does improve the performance for data practices involving the data shared with third-party, with 0.09 (+13.6\%) on precision and 0.08 (+13.1\%) on F1-score, respectively, but there is no noticeable increase for data practices involving the first-party data collected. Moreover, the fully LLM-based framework will result in approximately five times greater monetary cost.
\end{tcolorbox}
\section{Threats to Validity}
\label{sec_threats}

\noindent \textbf{Internal validity.} Threats to internal validity refer to errors in our experiments. To perform our experiments, we employ the power of GPT-3, a pre-trained LLM to automatically generate privacy nutrition labels from privacy policies. We follow OpenAI’s API to access GPT-3. We evaluate the performance of our framework on the Google Play app store dataset. We use \textit{langdetect} Python library to detect English texts in privacy policies. We reuse the trained multi-label classification model to identify privacy policy segments that are related to data practices. We have rechecked the code and data. However, there are still some remaining unknown errors. 

\noindent \textbf{External validity.} Threats to external validity refer to the generalizability of the study. In our experiments, we only employ applications from the Google Play app store and the Apple App Store. This may be a threat to external validity, as these applications may not be generalized beyond our investigations. As these stores are widely popular in mobile applications and require developers to provide high-quality privacy nutrition labels, we believe that there is a minimal threat to external validity. In the future, we plan to employ our framework on various datasets, such as websites and e-commerce, to reduce the threats to external validity.

\section{Related Work}
\label{sec_related_work}

Previous researchers have been investigating the idea of privacy nutrition labels from various perspectives. The iOS privacy nutrition label has been empirically studied, from its market-ready status~\cite{li2022understandingios} to how it is created~\cite{li2022understanding}. Emami-Naeini et al.~\cite{emami2019exploring, emami2021privacy} extensively studied the relationship between privacy attributes of IoT devices and the consumer's willingness to purchase, which showed privacy attributes as a major concern of consumers. They further reported the design space and experts' opinions on the content of the privacy label for IoT devices~\cite{emami2020ask}. Based on their findings, a prototype privacy label was proposed~\cite{emami2021informative}, aiming to inform consumers when making purchase decisions.

Another set of related works explored the potential of automatically generating privacy information.
Some of the work focuses on generating privacy policies from the source code of the programs, to achieve consistency between the behavior of the software and the corresponding privacy policies. Yu et al.~\cite{yu2015autoppg, TSE_2018} introduced a system called ``AutoPPG'' that scans the codebase of the software and generates privacy policies by comparing the function names and identifying the usage of privacy-related Android programming interfaces.
Another work by Zimmeck et al.~\cite{zimmeck2021privacyflash} leveraged code signatures to extract privacy attributes of iOS apps from their source code. 
Jain et al.~\cite{jain2021prigen} aimed to generate privacy policies using a machine translation model piped with static analysis on the permission-requiring code segments.
From the perspective of the language quality of generated privacy policies~\cite{ISTAS_2021}, they proposed a framework and compared the performance of LSTM and Bi-LSTM to train generative models for privacy policy generation.
Instead of focusing on privacy policies, Gardner et al.~\cite{gardner2022helping} identified data collection in the source code, and interactively assisted developers to provide precise privacy labels for their applications. 
Pan et al.~\cite{pan2023large} conducted a large-scale empirical study of developer-oriented online Automated Privacy Policy Generators (APPGs) for mobile apps.

Former works have unexplored the potential of automatically generating privacy labels from privacy policies. Our paper aims to fill the gap between the vastly available privacy policies and the newly required privacy nutrition labels from application markets, such as the Google Play app store and the Apple App Store. To the best of our knowledge, this framework is the first to comprehend privacy policies and privacy label formats to generate privacy nutrition labels, relieving customers' pain when they try to read the privacy policies. This approach has shown impressive performance and opened a new path for this research direction.
\section{Conclusion and Future Work}
\label{sec_conclusion}

The ``Privacy Policy Reading Phobia'' is widely recognized and discussed in previous studies. In this paper, we propose a framework that can automatically generate privacy nutrition labels from privacy policies, liberating software application users from hunting for information through privacy policies. Our framework achieves a 0.75 F1-score on generating \textit{first-party data collected}, a 0.63 F1-score on generating \textit{data shared with third-party}, and an average of 0.93 F1-score on general \textit{security practices} for Google's Data Safety Report.
We summarise three reasons undermining the performance of our framework, including omnibus data types, ambiguity caused by group-specific clauses, and under-performed context classification.
We also examine the inconsistencies between the ground truth and curated privacy nutrition labels. We find that there could be around 12.6\% data practices of the \textit{first-party data collected} are under-claimed and our framework can detect 90.1\% of under-claimed issues.
Our framework also demonstrates decent generalizability, yielding an average of 0.70 F1-score for Apple's App Privacy Details. 
In addition, we explore a fully LLM-based framework design, improving 13.1\% F1-score on generating \textit{data shared with third-party}, but no noticeable increasing for \textit{first-party data collected}. Also, the alternative design requires approximately five times more monetary cost. Lastly, we discuss the internal and external threats to the validity of our project. By providing customers with easy-to-understand summaries of privacy policies, companies can increase transparency and foster trust with their user base. This is particularly important in the context of data privacy, where consumers are increasingly concerned about the use and protection of their personal information.

Based on our promising results and findings, we recognize two potential directions for future research. First, we plan to implement this proposed framework as an online tool or a web browser extension, so customers can utilize this tool to generate the privacy nutrition label whenever they encounter a privacy policy. We also aim to improve and optimize the efficiency of our framework in the future. Second, we believe it is worthwhile to design a wide-recognized, user-friendly, and regulation-compliance standard privacy nutrition label format for websites. In addition, various privacy laws address relatively different and few personal data types. Therefore, a region-wise or regulation-wise privacy nutrition label format could be especially needed soon.
Above all, we believe our framework represents a step forward toward the cure of ``Privacy Policy Reading Phobia'', positively influencing users' behavior and promoting privacy-conscious decision-making.

\bibliographystyle{ACM-Reference-Format}
\bibliography{9_References}

%%% -*-BibTeX-*-
%%% Do NOT edit. File created by BibTeX with style
%%% ACM-Reference-Format-Journals [18-Jan-2012].

\begin{thebibliography}{90}

%%% ====================================================================
%%% NOTE TO THE USER: you can override these defaults by providing
%%% customized versions of any of these macros before the \bibliography
%%% command.  Each of them MUST provide its own final punctuation,
%%% except for \shownote{}, \showDOI{}, and \showURL{}.  The latter two
%%% do not use final punctuation, in order to avoid confusing it with
%%% the Web address.
%%%
%%% To suppress output of a particular field, define its macro to expand
%%% to an empty string, or better, \unskip, like this:
%%%
%%% \newcommand{\showDOI}[1]{\unskip}   % LaTeX syntax
%%%
%%% \def \showDOI #1{\unskip}           % plain TeX syntax
%%%
%%% ====================================================================

\ifx \showCODEN    \undefined \def \showCODEN     #1{\unskip}     \fi
\ifx \showDOI      \undefined \def \showDOI       #1{#1}\fi
\ifx \showISBNx    \undefined \def \showISBNx     #1{\unskip}     \fi
\ifx \showISBNxiii \undefined \def \showISBNxiii  #1{\unskip}     \fi
\ifx \showISSN     \undefined \def \showISSN      #1{\unskip}     \fi
\ifx \showLCCN     \undefined \def \showLCCN      #1{\unskip}     \fi
\ifx \shownote     \undefined \def \shownote      #1{#1}          \fi
\ifx \showarticletitle \undefined \def \showarticletitle #1{#1}   \fi
\ifx \showURL      \undefined \def \showURL       {\relax}        \fi
% The following commands are used for tagged output and should be
% invisible to TeX
\providecommand\bibfield[2]{#2}
\providecommand\bibinfo[2]{#2}
\providecommand\natexlab[1]{#1}
\providecommand\showeprint[2][]{arXiv:#2}

\bibitem[foo(2016)]%
        {foodlabel}
 \bibinfo{year}{2016}\natexlab{}.
\newblock \bibinfo{booktitle}{\emph{FDA Nutrition Facts Label}}.
\newblock
\urldef\tempurl%
\url{https://www.fda.gov/food/food-labeling-nutrition/changes-nutrition-facts-label}
\showURL{%
\tempurl}
\newblock
\shownote{Accessed: 2022-10-03}.


\bibitem[GDP(2016)]%
        {GDPR}
 \bibinfo{year}{2016}\natexlab{}.
\newblock \bibinfo{booktitle}{\emph{General Data Protection Regulation
  ({GDPR})}}.
\newblock
\urldef\tempurl%
\url{https://gdpr-info.eu/}
\showURL{%
\tempurl}
\newblock
\shownote{Accessed: 2022-04-25}.


\bibitem[CCP(2018)]%
        {CCPA}
 \bibinfo{year}{2018}\natexlab{}.
\newblock \bibinfo{booktitle}{\emph{California Consumer Privacy Act of 2018
  ({CCPA})}}.
\newblock
\urldef\tempurl%
\url{https://oag.ca.gov/privacy/ccpa}
\showURL{%
\tempurl}
\newblock
\shownote{Accessed: 2022-04-25}.


\bibitem[gpt(2022)]%
        {gpt3close}
 \bibinfo{year}{2022}\natexlab{}.
\newblock \bibinfo{booktitle}{\emph{Amplifying GPT-3 on closed-ended
  questions.}}
\newblock
\urldef\tempurl%
\url{https://generative.ink/posts/amplifying-gpt-3-on-closed-ended-questions/}
\showURL{%
\tempurl}
\newblock
\shownote{Accessed: 2022-10-03}.


\bibitem[app(2022a)]%
        {appleprivacy}
 \bibinfo{year}{2022}\natexlab{a}.
\newblock \bibinfo{booktitle}{\emph{App privacy details on the App Store}}.
\newblock
\urldef\tempurl%
\url{https://developer.apple.com/app-store/app-privacy-details/}
\showURL{%
\tempurl}
\newblock
\shownote{Accessed: 2022-10-10}.


\bibitem[app(2022b)]%
        {appbrain}
 \bibinfo{year}{2022}\natexlab{b}.
\newblock \bibinfo{booktitle}{\emph{AppBrain}}.
\newblock
\urldef\tempurl%
\url{https://www.appbrain.com/}
\showURL{%
\tempurl}
\newblock
\shownote{Accessed: 2022-08-26}.


\bibitem[bea(2022)]%
        {beautifulsoup}
 \bibinfo{year}{2022}\natexlab{}.
\newblock \bibinfo{booktitle}{\emph{BeautifulSoup}}.
\newblock
\urldef\tempurl%
\url{https://www.crummy.com/software/BeautifulSoup/}
\showURL{%
\tempurl}
\newblock
\shownote{Accessed: 2022-10-01}.


\bibitem[Iub(2022)]%
        {Iubenda}
 \bibinfo{year}{2022}\natexlab{}.
\newblock \bibinfo{booktitle}{\emph{Iubenda}}.
\newblock
\urldef\tempurl%
\url{https://www.iubenda.com/en/}
\showURL{%
\tempurl}
\newblock
\shownote{Accessed: 2022-03-28}.


\bibitem[lan(2022)]%
        {langdetect}
 \bibinfo{year}{2022}\natexlab{}.
\newblock \bibinfo{booktitle}{\emph{langdetect}}.
\newblock
\urldef\tempurl%
\url{https://pypi.org/project/langdetect/}
\showURL{%
\tempurl}
\newblock
\shownote{Accessed: 2022-10-01}.


\bibitem[Sel(2022)]%
        {Selenium}
 \bibinfo{year}{2022}\natexlab{}.
\newblock \bibinfo{booktitle}{\emph{Selenium}}.
\newblock
\urldef\tempurl%
\url{https://www.selenium.dev/}
\showURL{%
\tempurl}
\newblock
\shownote{Accessed: 2022-10-01}.


\bibitem[ter(2022)]%
        {termly}
 \bibinfo{year}{2022}\natexlab{}.
\newblock \bibinfo{booktitle}{\emph{Termly}}.
\newblock
\urldef\tempurl%
\url{https://termly.io/}
\showURL{%
\tempurl}
\newblock
\shownote{Accessed: 2022-03-28}.


\bibitem[goo(2022)]%
        {googledatasafety}
 \bibinfo{year}{2022}\natexlab{}.
\newblock \bibinfo{booktitle}{\emph{Understand app privacy and security
  practices with Google Play's Data safety section}}.
\newblock
\urldef\tempurl%
\url{https://support.google.com/googleplay/answer/11416267}
\showURL{%
\tempurl}
\newblock
\shownote{Accessed: 2022-10-10}.


\bibitem[web(2022)]%
        {websitepolicies}
 \bibinfo{year}{2022}\natexlab{}.
\newblock \bibinfo{booktitle}{\emph{Website Policies}}.
\newblock
\urldef\tempurl%
\url{https://www.websitepolicies.com/}
\showURL{%
\tempurl}
\newblock
\shownote{Accessed: 2022-03-28}.


\bibitem[Goo(2023)]%
        {GoogleBard}
 \bibinfo{year}{2023}\natexlab{}.
\newblock \bibinfo{booktitle}{\emph{Bard}}.
\newblock
\urldef\tempurl%
\url{https://bard.google.com/}
\showURL{%
\tempurl}
\newblock
\shownote{Accessed: 2022-03-27}.


\bibitem[cha(2023)]%
        {chatGPT}
 \bibinfo{year}{2023}\natexlab{}.
\newblock \bibinfo{booktitle}{\emph{chatGPT}}.
\newblock
\urldef\tempurl%
\url{https://openai.com/blog/chatgpt}
\showURL{%
\tempurl}
\newblock
\shownote{Accessed: 2022-03-27}.


\bibitem[Mat(2023)]%
        {MatchMasters}
 \bibinfo{year}{2023}\natexlab{}.
\newblock \bibinfo{booktitle}{\emph{Match Masters}}.
\newblock
\urldef\tempurl%
\url{https://play.google.com/store/apps/details?id=com.funtomic.matchmasters}
\showURL{%
\tempurl}
\newblock
\shownote{Accessed: 2023-03-24}.


\bibitem[See(2023)]%
        {SeeNoEvil}
 \bibinfo{year}{2023}\natexlab{}.
\newblock \bibinfo{booktitle}{\emph{See No Evil: Loopholes in Google’s Data
  Safety Labels Keep Companies in the Clear and Consumers in the Dark}}.
\newblock
\urldef\tempurl%
\url{https://foundation.mozilla.org/en/campaigns/googles-data-safety-labels/}
\showURL{%
\tempurl}
\newblock
\shownote{Accessed: 2022-03-27}.


\bibitem[Sup(2023)]%
        {SuperSlimeSimulator}
 \bibinfo{year}{2023}\natexlab{}.
\newblock \bibinfo{booktitle}{\emph{Super Slime Simulator: DIY Art}}.
\newblock
\urldef\tempurl%
\url{https://play.google.com/store/apps/details?id=com.dramaton.slime}
\showURL{%
\tempurl}
\newblock
\shownote{Accessed: 2023-03-24}.


\bibitem[Allamanis et~al\mbox{.}(2021)]%
        {allamanis2021self}
\bibfield{author}{\bibinfo{person}{Miltiadis Allamanis}, \bibinfo{person}{Henry
  Jackson-Flux}, {and} \bibinfo{person}{Marc Brockschmidt}.}
  \bibinfo{year}{2021}\natexlab{}.
\newblock \showarticletitle{Self-supervised bug detection and repair}.
\newblock \bibinfo{journal}{\emph{Advances in Neural Information Processing
  Systems}}  \bibinfo{volume}{34} (\bibinfo{year}{2021}),
  \bibinfo{pages}{27865--27876}.
\newblock


\bibitem[Allix et~al\mbox{.}(2016)]%
        {androzoo}
\bibfield{author}{\bibinfo{person}{Kevin Allix},
  \bibinfo{person}{Tegawend{\'e}~F. Bissyand{\'e}}, \bibinfo{person}{Jacques
  Klein}, {and} \bibinfo{person}{Yves Le~Traon}.}
  \bibinfo{year}{2016}\natexlab{}.
\newblock \showarticletitle{AndroZoo: Collecting Millions of Android Apps for
  the Research Community}. In \bibinfo{booktitle}{\emph{Proceedings of the 13th
  International Conference on Mining Software Repositories}} (Austin, Texas)
  \emph{(\bibinfo{series}{MSR '16})}. \bibinfo{publisher}{ACM},
  \bibinfo{address}{New York, NY, USA}, \bibinfo{pages}{468--471}.
\newblock
\showISBNx{978-1-4503-4186-8}
\urldef\tempurl%
\url{https://doi.org/10.1145/2901739.2903508}
\showDOI{\tempurl}


\bibitem[Apolinarski et~al\mbox{.}(2015)]%
        {apolinarski2015automating}
\bibfield{author}{\bibinfo{person}{Wolfgang Apolinarski},
  \bibinfo{person}{Marcus Handte}, {and} \bibinfo{person}{Pedro~Jose Marron}.}
  \bibinfo{year}{2015}\natexlab{}.
\newblock \showarticletitle{Automating the generation of privacy policies for
  context-sharing applications}. In \bibinfo{booktitle}{\emph{2015
  International Conference on Intelligent Environments}}. IEEE,
  \bibinfo{pages}{73--80}.
\newblock


\bibitem[Balebako et~al\mbox{.}(2014)]%
        {balebako2014your}
\bibfield{author}{\bibinfo{person}{Rebecca Balebako}, \bibinfo{person}{Richard
  Shay}, {and} \bibinfo{person}{Lorrie~Faith Cranor}.}
  \bibinfo{year}{2014}\natexlab{}.
\newblock \showarticletitle{Is your inseam a biometric? a case study on the
  role of usability studies in developing public policy}.
\newblock \bibinfo{journal}{\emph{Proc. USEC}}  \bibinfo{volume}{14}
  (\bibinfo{year}{2014}).
\newblock


\bibitem[Barth and De~Jong(2017)]%
        {barth2017privacy}
\bibfield{author}{\bibinfo{person}{Susanne Barth} {and}
  \bibinfo{person}{Menno~DT De~Jong}.} \bibinfo{year}{2017}\natexlab{}.
\newblock \showarticletitle{The privacy paradox--Investigating discrepancies
  between expressed privacy concerns and actual online behavior--A systematic
  literature review}.
\newblock \bibinfo{journal}{\emph{Telematics and informatics}}
  \bibinfo{volume}{34}, \bibinfo{number}{7} (\bibinfo{year}{2017}),
  \bibinfo{pages}{1038--1058}.
\newblock


\bibitem[Bateni and Dara(2021)]%
        {ISTAS_2021}
\bibfield{author}{\bibinfo{person}{Nastaran Bateni} {and}
  \bibinfo{person}{Rozita Dara}.} \bibinfo{year}{2021}\natexlab{}.
\newblock \showarticletitle{Automated generation of privacy policy using deep
  models}. In \bibinfo{booktitle}{\emph{2021 IEEE International Symposium on
  Technology and Society (ISTAS)}}. IEEE, \bibinfo{pages}{1--6}.
\newblock


\bibitem[Blakkarly and Graham(2022)]%
        {Blakkarly2022privacy}
\bibfield{author}{\bibinfo{person}{Jarni Blakkarly} {and}
  \bibinfo{person}{Daniel Graham}.} \bibinfo{year}{2022}\natexlab{}.
\newblock \bibinfo{booktitle}{\emph{Privacy policy comparison reveals half have
  poor readability}}.
\newblock
\urldef\tempurl%
\url{https://www.choice.com.au/consumers-and-data/protecting-your-data/data-laws-and-regulation/articles/privacy-policy-comparison}
\showURL{%
\tempurl}
\newblock
\shownote{Accessed: 2022-05-03}.


\bibitem[Bojanowski et~al\mbox{.}(2016)]%
        {bojanowski2016enriching}
\bibfield{author}{\bibinfo{person}{Piotr Bojanowski}, \bibinfo{person}{Edouard
  Grave}, \bibinfo{person}{Armand Joulin}, {and} \bibinfo{person}{Tomas
  Mikolov}.} \bibinfo{year}{2016}\natexlab{}.
\newblock \showarticletitle{Enriching Word Vectors with Subword Information}.
\newblock \bibinfo{journal}{\emph{arXiv preprint arXiv:1607.04606}}
  (\bibinfo{year}{2016}).
\newblock


\bibitem[Brown et~al\mbox{.}(2020)]%
        {brown2020language}
\bibfield{author}{\bibinfo{person}{Tom Brown}, \bibinfo{person}{Benjamin Mann},
  \bibinfo{person}{Nick Ryder}, \bibinfo{person}{Melanie Subbiah},
  \bibinfo{person}{Jared~D Kaplan}, \bibinfo{person}{Prafulla Dhariwal},
  \bibinfo{person}{Arvind Neelakantan}, \bibinfo{person}{Pranav Shyam},
  \bibinfo{person}{Girish Sastry}, \bibinfo{person}{Amanda Askell},
  {et~al\mbox{.}}} \bibinfo{year}{2020}\natexlab{}.
\newblock \showarticletitle{Language models are few-shot learners}.
\newblock \bibinfo{journal}{\emph{Advances in neural information processing
  systems}}  \bibinfo{volume}{33} (\bibinfo{year}{2020}),
  \bibinfo{pages}{1877--1901}.
\newblock


\bibitem[Caramujo and Da~Silva(2015)]%
        {caramujo2015analyzing}
\bibfield{author}{\bibinfo{person}{Jo{\~a}o Caramujo} {and}
  \bibinfo{person}{Alberto Manuel~Rodrigues Da~Silva}.}
  \bibinfo{year}{2015}\natexlab{}.
\newblock \showarticletitle{Analyzing privacy policies based on a privacy-aware
  profile: The Facebook and LinkedIn case studies}. In
  \bibinfo{booktitle}{\emph{2015 IEEE 17th Conference on Business
  Informatics}}, Vol.~\bibinfo{volume}{1}. IEEE, \bibinfo{pages}{77--84}.
\newblock


\bibitem[Ciocchetti(2008)]%
        {ciocchetti2008future}
\bibfield{author}{\bibinfo{person}{Corey~A Ciocchetti}.}
  \bibinfo{year}{2008}\natexlab{}.
\newblock \showarticletitle{The future of privacy policies: A privacy nutrition
  label filled with fair information practices}.
\newblock \bibinfo{journal}{\emph{J. Marshall J. Computer \& Info. L.}}
  \bibinfo{volume}{26} (\bibinfo{year}{2008}), \bibinfo{pages}{1}.
\newblock


\bibitem[Cranor(2012)]%
        {cranor2012necessary}
\bibfield{author}{\bibinfo{person}{Lorrie~Faith Cranor}.}
  \bibinfo{year}{2012}\natexlab{}.
\newblock \showarticletitle{Necessary but not sufficient: Standardized
  mechanisms for privacy notice and choice}.
\newblock \bibinfo{journal}{\emph{J. on Telecomm. \& High Tech. L.}}
  \bibinfo{volume}{10} (\bibinfo{year}{2012}), \bibinfo{pages}{273}.
\newblock


\bibitem[Datta et~al\mbox{.}(2014)]%
        {datta2014automated}
\bibfield{author}{\bibinfo{person}{Amit Datta}, \bibinfo{person}{Michael~Carl
  Tschantz}, {and} \bibinfo{person}{Anupam Datta}.}
  \bibinfo{year}{2014}\natexlab{}.
\newblock \showarticletitle{Automated experiments on ad privacy settings: A
  tale of opacity, choice, and discrimination}.
\newblock \bibinfo{journal}{\emph{arXiv preprint arXiv:1408.6491}}
  (\bibinfo{year}{2014}).
\newblock


\bibitem[Dommeyer and Gross(2003)]%
        {dommeyer2003consumers}
\bibfield{author}{\bibinfo{person}{Curt~J Dommeyer} {and}
  \bibinfo{person}{Barbara~L Gross}.} \bibinfo{year}{2003}\natexlab{}.
\newblock \showarticletitle{What consumers know and what they do: An
  investigation of consumer knowledge, awareness, and use of privacy protection
  strategies}.
\newblock \bibinfo{journal}{\emph{Journal of Interactive Marketing}}
  \bibinfo{volume}{17}, \bibinfo{number}{2} (\bibinfo{year}{2003}),
  \bibinfo{pages}{34--51}.
\newblock


\bibitem[Ellawela and Lakmali(2021)]%
        {ellawela2021review}
\bibfield{author}{\bibinfo{person}{Chaveen Ellawela} {and} \bibinfo{person}{KBN
  Lakmali}.} \bibinfo{year}{2021}\natexlab{}.
\newblock \showarticletitle{A Review about Voice and UI Design Driven
  Approaches to Identify UI Elements and Generate UI Designs}. In
  \bibinfo{booktitle}{\emph{2021 International Conference on Intelligent
  Technologies (CONIT)}}. IEEE, \bibinfo{pages}{1--4}.
\newblock


\bibitem[Emami-Naeini et~al\mbox{.}(2020)]%
        {emami2020ask}
\bibfield{author}{\bibinfo{person}{Pardis Emami-Naeini},
  \bibinfo{person}{Yuvraj Agarwal}, \bibinfo{person}{Lorrie~Faith Cranor},
  {and} \bibinfo{person}{Hanan Hibshi}.} \bibinfo{year}{2020}\natexlab{}.
\newblock \showarticletitle{Ask the experts: What should be on an IoT privacy
  and security label?}. In \bibinfo{booktitle}{\emph{2020 IEEE Symposium on
  Security and Privacy (SP)}}. IEEE, \bibinfo{pages}{447--464}.
\newblock


\bibitem[Emami-Naeini et~al\mbox{.}(2021a)]%
        {emami2021informative}
\bibfield{author}{\bibinfo{person}{Pardis Emami-Naeini},
  \bibinfo{person}{Janarth Dheenadhayalan}, \bibinfo{person}{Yuvraj Agarwal},
  {and} \bibinfo{person}{Lorrie~Faith Cranor}.}
  \bibinfo{year}{2021}\natexlab{a}.
\newblock \showarticletitle{An Informative Security and Privacy “Nutrition”
  Label for Internet of Things Devices}.
\newblock \bibinfo{journal}{\emph{IEEE Security \& Privacy}}
  \bibinfo{volume}{20}, \bibinfo{number}{2} (\bibinfo{year}{2021}),
  \bibinfo{pages}{31--39}.
\newblock


\bibitem[Emami-Naeini et~al\mbox{.}(2021b)]%
        {emami2021privacy}
\bibfield{author}{\bibinfo{person}{Pardis Emami-Naeini},
  \bibinfo{person}{Janarth Dheenadhayalan}, \bibinfo{person}{Yuvraj Agarwal},
  {and} \bibinfo{person}{Lorrie~Faith Cranor}.}
  \bibinfo{year}{2021}\natexlab{b}.
\newblock \showarticletitle{Which privacy and security attributes most impact
  consumers’ risk perception and willingness to purchase IoT devices?}. In
  \bibinfo{booktitle}{\emph{2021 IEEE Symposium on Security and Privacy (SP)}}.
  IEEE, \bibinfo{pages}{519--536}.
\newblock


\bibitem[Emami-Naeini et~al\mbox{.}(2019)]%
        {emami2019exploring}
\bibfield{author}{\bibinfo{person}{Pardis Emami-Naeini}, \bibinfo{person}{Henry
  Dixon}, \bibinfo{person}{Yuvraj Agarwal}, {and} \bibinfo{person}{Lorrie~Faith
  Cranor}.} \bibinfo{year}{2019}\natexlab{}.
\newblock \showarticletitle{Exploring how privacy and security factor into IoT
  device purchase behavior}. In \bibinfo{booktitle}{\emph{Proceedings of the
  2019 CHI Conference on Human Factors in Computing Systems}}.
  \bibinfo{pages}{1--12}.
\newblock


\bibitem[Floridi and Chiriatti(2020)]%
        {floridi2020gpt}
\bibfield{author}{\bibinfo{person}{Luciano Floridi} {and}
  \bibinfo{person}{Massimo Chiriatti}.} \bibinfo{year}{2020}\natexlab{}.
\newblock \showarticletitle{GPT-3: Its nature, scope, limits, and
  consequences}.
\newblock \bibinfo{journal}{\emph{Minds and Machines}} \bibinfo{volume}{30},
  \bibinfo{number}{4} (\bibinfo{year}{2020}), \bibinfo{pages}{681--694}.
\newblock


\bibitem[Gardner et~al\mbox{.}(2022)]%
        {gardner2022helping}
\bibfield{author}{\bibinfo{person}{Jack Gardner}, \bibinfo{person}{Yuanyuan
  Feng}, \bibinfo{person}{Kayla Reiman}, \bibinfo{person}{Zhi Lin},
  \bibinfo{person}{Akshath Jain}, {and} \bibinfo{person}{Norman Sadeh}.}
  \bibinfo{year}{2022}\natexlab{}.
\newblock \showarticletitle{Helping mobile application developers create
  accurate privacy labels}. In \bibinfo{booktitle}{\emph{2022 IEEE European
  Symposium on Security and Privacy Workshops (EuroS\&PW)}}. IEEE,
  \bibinfo{pages}{212--230}.
\newblock


\bibitem[Gulcehre et~al\mbox{.}(2017)]%
        {gulcehre2017integrating}
\bibfield{author}{\bibinfo{person}{Caglar Gulcehre}, \bibinfo{person}{Orhan
  Firat}, \bibinfo{person}{Kelvin Xu}, \bibinfo{person}{Kyunghyun Cho}, {and}
  \bibinfo{person}{Yoshua Bengio}.} \bibinfo{year}{2017}\natexlab{}.
\newblock \showarticletitle{On integrating a language model into neural machine
  translation}.
\newblock \bibinfo{journal}{\emph{Computer Speech \& Language}}
  \bibinfo{volume}{45} (\bibinfo{year}{2017}), \bibinfo{pages}{137--148}.
\newblock


\bibitem[Harkous et~al\mbox{.}(2018)]%
        {harkous2018polisis}
\bibfield{author}{\bibinfo{person}{Hamza Harkous}, \bibinfo{person}{Kassem
  Fawaz}, \bibinfo{person}{R{\'e}mi Lebret}, \bibinfo{person}{Florian Schaub},
  \bibinfo{person}{Kang~G Shin}, {and} \bibinfo{person}{Karl Aberer}.}
  \bibinfo{year}{2018}\natexlab{}.
\newblock \showarticletitle{Polisis: Automated analysis and presentation of
  privacy policies using deep learning}. In \bibinfo{booktitle}{\emph{27th
  USENIX Security Symposium (USENIX Security 18)}}. \bibinfo{pages}{531--548}.
\newblock


\bibitem[Hendrycks et~al\mbox{.}(2020)]%
        {hendrycks2020measuring}
\bibfield{author}{\bibinfo{person}{Dan Hendrycks}, \bibinfo{person}{Collin
  Burns}, \bibinfo{person}{Steven Basart}, \bibinfo{person}{Andy Zou},
  \bibinfo{person}{Mantas Mazeika}, \bibinfo{person}{Dawn Song}, {and}
  \bibinfo{person}{Jacob Steinhardt}.} \bibinfo{year}{2020}\natexlab{}.
\newblock \showarticletitle{Measuring massive multitask language
  understanding}.
\newblock \bibinfo{journal}{\emph{arXiv preprint arXiv:2009.03300}}
  (\bibinfo{year}{2020}).
\newblock


\bibitem[Jain et~al\mbox{.}(2021)]%
        {jain2021prigen}
\bibfield{author}{\bibinfo{person}{Vijayanta Jain},
  \bibinfo{person}{Sanonda~Datta Gupta}, \bibinfo{person}{Sepideh Ghanavati},
  {and} \bibinfo{person}{Sai~Teja Peddinti}.} \bibinfo{year}{2021}\natexlab{}.
\newblock \showarticletitle{Prigen: Towards automated translation of android
  applications’ code to privacy captions}. In
  \bibinfo{booktitle}{\emph{International Conference on Research Challenges in
  Information Science}}. Springer, \bibinfo{pages}{142--151}.
\newblock


\bibitem[Kelley et~al\mbox{.}(2009)]%
        {kelley2009nutrition}
\bibfield{author}{\bibinfo{person}{Patrick~Gage Kelley},
  \bibinfo{person}{Joanna Bresee}, \bibinfo{person}{Lorrie~Faith Cranor}, {and}
  \bibinfo{person}{Robert~W Reeder}.} \bibinfo{year}{2009}\natexlab{}.
\newblock \showarticletitle{A" nutrition label" for privacy}. In
  \bibinfo{booktitle}{\emph{Proceedings of the 5th Symposium on Usable Privacy
  and Security}}. \bibinfo{pages}{1--12}.
\newblock


\bibitem[Kelley et~al\mbox{.}(2010)]%
        {kelley2010standardizing}
\bibfield{author}{\bibinfo{person}{Patrick~Gage Kelley},
  \bibinfo{person}{Lucian Cesca}, \bibinfo{person}{Joanna Bresee}, {and}
  \bibinfo{person}{Lorrie~Faith Cranor}.} \bibinfo{year}{2010}\natexlab{}.
\newblock \showarticletitle{Standardizing privacy notices: an online study of
  the nutrition label approach}. In \bibinfo{booktitle}{\emph{Proceedings of
  the SIGCHI Conference on Human factors in Computing Systems}}.
  \bibinfo{pages}{1573--1582}.
\newblock


\bibitem[Kelley et~al\mbox{.}(2013)]%
        {kelley2013privacy}
\bibfield{author}{\bibinfo{person}{Patrick~Gage Kelley},
  \bibinfo{person}{Lorrie~Faith Cranor}, {and} \bibinfo{person}{Norman Sadeh}.}
  \bibinfo{year}{2013}\natexlab{}.
\newblock \showarticletitle{Privacy as part of the app decision-making
  process}. In \bibinfo{booktitle}{\emph{Proceedings of the SIGCHI conference
  on human factors in computing systems}}. \bibinfo{pages}{3393--3402}.
\newblock


\bibitem[Kemp(2020)]%
        {kemp2020concealed}
\bibfield{author}{\bibinfo{person}{Katharine Kemp}.}
  \bibinfo{year}{2020}\natexlab{}.
\newblock \showarticletitle{Concealed data practices and competition law: why
  privacy matters}.
\newblock \bibinfo{journal}{\emph{European Competition Journal}}
  \bibinfo{volume}{16}, \bibinfo{number}{2-3} (\bibinfo{year}{2020}),
  \bibinfo{pages}{628--672}.
\newblock


\bibitem[Kincaid et~al\mbox{.}(1975)]%
        {kincaid1975derivation}
\bibfield{author}{\bibinfo{person}{J~Peter Kincaid}, \bibinfo{person}{Robert~P
  Fishburne~Jr}, \bibinfo{person}{Richard~L Rogers}, {and}
  \bibinfo{person}{Brad~S Chissom}.} \bibinfo{year}{1975}\natexlab{}.
\newblock \bibinfo{booktitle}{\emph{Derivation of new readability formulas
  (automated readability index, fog count and flesch reading ease formula) for
  navy enlisted personnel}}.
\newblock \bibinfo{type}{{T}echnical {R}eport}. \bibinfo{institution}{Naval
  Technical Training Command Millington TN Research Branch}.
\newblock


\bibitem[Kojima et~al\mbox{.}(2022)]%
        {kojima2022large}
\bibfield{author}{\bibinfo{person}{Takeshi Kojima},
  \bibinfo{person}{Shixiang~Shane Gu}, \bibinfo{person}{Machel Reid},
  \bibinfo{person}{Yutaka Matsuo}, {and} \bibinfo{person}{Yusuke Iwasawa}.}
  \bibinfo{year}{2022}\natexlab{}.
\newblock \showarticletitle{Large Language Models are Zero-Shot Reasoners}.
\newblock \bibinfo{journal}{\emph{arXiv preprint arXiv:2205.11916}}
  (\bibinfo{year}{2022}).
\newblock


\bibitem[Krumay and Klar(2020)]%
        {krumay2020readability}
\bibfield{author}{\bibinfo{person}{Barbara Krumay} {and}
  \bibinfo{person}{Jennifer Klar}.} \bibinfo{year}{2020}\natexlab{}.
\newblock \showarticletitle{Readability of privacy policies}. In
  \bibinfo{booktitle}{\emph{IFIP Annual Conference on Data and Applications
  Security and Privacy}}. Springer, \bibinfo{pages}{388--399}.
\newblock


\bibitem[Kumar et~al\mbox{.}(2022)]%
        {USENIX_2022_GEODIFF}
\bibfield{author}{\bibinfo{person}{Renuka Kumar}, \bibinfo{person}{Apurva
  Virkud}, \bibinfo{person}{Ram Sundara~Raman}, \bibinfo{person}{Atul Prakash},
  {and} \bibinfo{person}{Roya Ensafi}.} \bibinfo{year}{2022}\natexlab{}.
\newblock \showarticletitle{A Large-scale Investigation into Geodifferences in
  Mobile Apps}. In \bibinfo{booktitle}{\emph{31st USENIX Security Symposium
  (USENIX Security 22)}}. \bibinfo{publisher}{USENIX Association},
  \bibinfo{address}{Boston, MA}.
\newblock
\urldef\tempurl%
\url{https://www.usenix.org/conference/usenixsecurity22/presentation/kumar}
\showURL{%
\tempurl}


\bibitem[Lewicka(1998)]%
        {lewicka1998confirmation}
\bibfield{author}{\bibinfo{person}{Maria Lewicka}.}
  \bibinfo{year}{1998}\natexlab{}.
\newblock \showarticletitle{Confirmation bias}.
\newblock In \bibinfo{booktitle}{\emph{Personal control in action}}.
  \bibinfo{publisher}{Springer}, \bibinfo{pages}{233--258}.
\newblock


\bibitem[Li et~al\mbox{.}(2022b)]%
        {li2022understanding}
\bibfield{author}{\bibinfo{person}{Tianshi Li}, \bibinfo{person}{Kayla Reiman},
  \bibinfo{person}{Yuvraj Agarwal}, \bibinfo{person}{Lorrie~Faith Cranor},
  {and} \bibinfo{person}{Jason~I Hong}.} \bibinfo{year}{2022}\natexlab{b}.
\newblock \showarticletitle{Understanding Challenges for Developers to Create
  Accurate Privacy Nutrition Labels}. In \bibinfo{booktitle}{\emph{CHI
  Conference on Human Factors in Computing Systems}}. \bibinfo{pages}{1--24}.
\newblock


\bibitem[Li et~al\mbox{.}(2022a)]%
        {li2022understandingios}
\bibfield{author}{\bibinfo{person}{Yucheng Li}, \bibinfo{person}{Deyuan Chen},
  \bibinfo{person}{Tianshi Li}, \bibinfo{person}{Yuvraj Agarwal},
  \bibinfo{person}{Lorrie~Faith Cranor}, {and} \bibinfo{person}{Jason~I Hong}.}
  \bibinfo{year}{2022}\natexlab{a}.
\newblock \showarticletitle{Understanding iOS Privacy Nutrition Labels: An
  Exploratory Large-Scale Analysis of App Store Data}. In
  \bibinfo{booktitle}{\emph{CHI Conference on Human Factors in Computing
  Systems Extended Abstracts}}. \bibinfo{pages}{1--7}.
\newblock


\bibitem[Liu et~al\mbox{.}(2014)]%
        {liu2014step}
\bibfield{author}{\bibinfo{person}{Fei Liu}, \bibinfo{person}{Rohan Ramanath},
  \bibinfo{person}{Norman Sadeh}, {and} \bibinfo{person}{Noah~A Smith}.}
  \bibinfo{year}{2014}\natexlab{}.
\newblock \showarticletitle{A step towards usable privacy policy: Automatic
  alignment of privacy statements}. In \bibinfo{booktitle}{\emph{Proceedings of
  COLING 2014, the 25th International Conference on Computational Linguistics:
  Technical Papers}}. \bibinfo{pages}{884--894}.
\newblock


\bibitem[Liu et~al\mbox{.}(2018)]%
        {CMU_2017}
\bibfield{author}{\bibinfo{person}{Frederick Liu}, \bibinfo{person}{Shomir
  Wilson}, \bibinfo{person}{Peter Story}, \bibinfo{person}{Sebastian Zimmeck},
  {and} \bibinfo{person}{Norman Sadeh}.} \bibinfo{year}{2018}\natexlab{}.
\newblock \showarticletitle{Towards automatic classification of privacy policy
  text}.
\newblock \bibinfo{journal}{\emph{School of Computer Science Carnegie Mellon
  University}} (\bibinfo{year}{2018}).
\newblock


\bibitem[Liu et~al\mbox{.}(2021)]%
        {liu2021have}
\bibfield{author}{\bibinfo{person}{Shuang Liu}, \bibinfo{person}{Baiyang Zhao},
  \bibinfo{person}{Renjie Guo}, \bibinfo{person}{Guozhu Meng},
  \bibinfo{person}{Fan Zhang}, {and} \bibinfo{person}{Meishan Zhang}.}
  \bibinfo{year}{2021}\natexlab{}.
\newblock \showarticletitle{Have You been Properly Notified? Automatic
  Compliance Analysis of Privacy Policy Text with GDPR Article 13}. In
  \bibinfo{booktitle}{\emph{Proceedings of the Web Conference 2021}}.
  \bibinfo{pages}{2154--2164}.
\newblock


\bibitem[Mercier(2016)]%
        {mercier2016confirmation}
\bibfield{author}{\bibinfo{person}{Hugo Mercier}.}
  \bibinfo{year}{2016}\natexlab{}.
\newblock \showarticletitle{Confirmation bias--myside bias}.
\newblock In \bibinfo{booktitle}{\emph{Cognitive illusions}}.
  \bibinfo{publisher}{Psychology Press}, \bibinfo{pages}{109--124}.
\newblock


\bibitem[Mialon et~al\mbox{.}(2023)]%
        {mialon2023augmented}
\bibfield{author}{\bibinfo{person}{Gr{\'e}goire Mialon},
  \bibinfo{person}{Roberto Dess{\`\i}}, \bibinfo{person}{Maria Lomeli},
  \bibinfo{person}{Christoforos Nalmpantis}, \bibinfo{person}{Ram Pasunuru},
  \bibinfo{person}{Roberta Raileanu}, \bibinfo{person}{Baptiste Rozi{\`e}re},
  \bibinfo{person}{Timo Schick}, \bibinfo{person}{Jane Dwivedi-Yu},
  \bibinfo{person}{Asli Celikyilmaz}, {et~al\mbox{.}}}
  \bibinfo{year}{2023}\natexlab{}.
\newblock \showarticletitle{Augmented language models: a survey}.
\newblock \bibinfo{journal}{\emph{arXiv preprint arXiv:2302.07842}}
  (\bibinfo{year}{2023}).
\newblock


\bibitem[Mikians et~al\mbox{.}(2012)]%
        {mikians2012detecting}
\bibfield{author}{\bibinfo{person}{Jakub Mikians},
  \bibinfo{person}{L{\'a}szl{\'o} Gyarmati}, \bibinfo{person}{Vijay Erramilli},
  {and} \bibinfo{person}{Nikolaos Laoutaris}.} \bibinfo{year}{2012}\natexlab{}.
\newblock \showarticletitle{Detecting price and search discrimination on the
  internet}. In \bibinfo{booktitle}{\emph{Proceedings of the 11th ACM workshop
  on hot topics in networks}}. \bibinfo{pages}{79--84}.
\newblock


\bibitem[Min et~al\mbox{.}(2021)]%
        {min2021recent}
\bibfield{author}{\bibinfo{person}{Bonan Min}, \bibinfo{person}{Hayley Ross},
  \bibinfo{person}{Elior Sulem}, \bibinfo{person}{Amir Pouran~Ben Veyseh},
  \bibinfo{person}{Thien~Huu Nguyen}, \bibinfo{person}{Oscar Sainz},
  \bibinfo{person}{Eneko Agirre}, \bibinfo{person}{Ilana Heinz}, {and}
  \bibinfo{person}{Dan Roth}.} \bibinfo{year}{2021}\natexlab{}.
\newblock \showarticletitle{Recent advances in natural language processing via
  large pre-trained language models: A survey}.
\newblock \bibinfo{journal}{\emph{arXiv preprint arXiv:2111.01243}}
  (\bibinfo{year}{2021}).
\newblock


\bibitem[Mohan et~al\mbox{.}(2019)]%
        {mohan2019analyzing}
\bibfield{author}{\bibinfo{person}{Jayashree Mohan}, \bibinfo{person}{Melissa
  Wasserman}, {and} \bibinfo{person}{Vijay Chidambaram}.}
  \bibinfo{year}{2019}\natexlab{}.
\newblock \showarticletitle{Analyzing GDPR compliance through the lens of
  privacy policy}. In \bibinfo{booktitle}{\emph{Heterogeneous Data Management,
  Polystores, and Analytics for Healthcare: VLDB 2019 Workshops, Poly and DMAH,
  Los Angeles, CA, USA, August 30, 2019, Revised Selected Papers 5}}. Springer,
  \bibinfo{pages}{82--95}.
\newblock


\bibitem[Mulder(2019)]%
        {mulder2019health}
\bibfield{author}{\bibinfo{person}{Trix Mulder}.}
  \bibinfo{year}{2019}\natexlab{}.
\newblock \showarticletitle{Health apps, their privacy policies and the GDPR}.
\newblock \bibinfo{journal}{\emph{European Journal of Law and Technology}}
  (\bibinfo{year}{2019}).
\newblock


\bibitem[Narasimhan et~al\mbox{.}(2021)]%
        {narasimhan2021cgems}
\bibfield{author}{\bibinfo{person}{Aishwarya Narasimhan},
  \bibinfo{person}{Krishna Prasad Agara~Venkatesha Rao}, {et~al\mbox{.}}}
  \bibinfo{year}{2021}\natexlab{}.
\newblock \showarticletitle{CGEMs: A Metric Model for Automatic Code Generation
  using GPT-3}.
\newblock \bibinfo{journal}{\emph{arXiv preprint arXiv:2108.10168}}
  (\bibinfo{year}{2021}).
\newblock


\bibitem[Nickerson(1998)]%
        {nickerson1998confirmation}
\bibfield{author}{\bibinfo{person}{Raymond~S Nickerson}.}
  \bibinfo{year}{1998}\natexlab{}.
\newblock \showarticletitle{Confirmation bias: A ubiquitous phenomenon in many
  guises}.
\newblock \bibinfo{journal}{\emph{Review of general psychology}}
  \bibinfo{volume}{2}, \bibinfo{number}{2} (\bibinfo{year}{1998}),
  \bibinfo{pages}{175--220}.
\newblock


\bibitem[Obar and Oeldorf-Hirsch(2020)]%
        {obar2020biggest}
\bibfield{author}{\bibinfo{person}{Jonathan~A Obar} {and} \bibinfo{person}{Anne
  Oeldorf-Hirsch}.} \bibinfo{year}{2020}\natexlab{}.
\newblock \showarticletitle{The biggest lie on the internet: Ignoring the
  privacy policies and terms of service policies of social networking
  services}.
\newblock \bibinfo{journal}{\emph{Information, Communication \& Society}}
  \bibinfo{volume}{23}, \bibinfo{number}{1} (\bibinfo{year}{2020}),
  \bibinfo{pages}{128--147}.
\newblock


\bibitem[OpenAI(2023)]%
        {openai2023gpt4}
\bibfield{author}{\bibinfo{person}{OpenAI}.} \bibinfo{year}{2023}\natexlab{}.
\newblock \bibinfo{title}{GPT-4 Technical Report}.
\newblock
\newblock
\showeprint[arxiv]{2303.08774}~[cs.CL]


\bibitem[Paik and Wang(2021)]%
        {paik2021improving}
\bibfield{author}{\bibinfo{person}{Incheon Paik} {and} \bibinfo{person}{Jun-Wei
  Wang}.} \bibinfo{year}{2021}\natexlab{}.
\newblock \showarticletitle{Improving Text-to-Code Generation with Features of
  Code Graph on GPT-2}.
\newblock \bibinfo{journal}{\emph{Electronics}} \bibinfo{volume}{10},
  \bibinfo{number}{21} (\bibinfo{year}{2021}), \bibinfo{pages}{2706}.
\newblock


\bibitem[Pan et~al\mbox{.}(2023)]%
        {pan2023large}
\bibfield{author}{\bibinfo{person}{Shidong Pan}, \bibinfo{person}{Dawen Zhang},
  \bibinfo{person}{Mark Staples}, \bibinfo{person}{Zhenchang Xing},
  \bibinfo{person}{Jieshan Chen}, \bibinfo{person}{Xiwei Xu}, {and}
  \bibinfo{person}{James Hoang}.} \bibinfo{year}{2023}\natexlab{}.
\newblock \showarticletitle{A Large-scale Empirical Study of Online Automated
  Privacy Policy Generators for Mobile Apps}.
\newblock \bibinfo{journal}{\emph{arXiv preprint arXiv:2305.03271}}
  (\bibinfo{year}{2023}).
\newblock


\bibitem[Perez et~al\mbox{.}(2018)]%
        {perez2018review}
\bibfield{author}{\bibinfo{person}{Alfredo~J Perez}, \bibinfo{person}{Sherali
  Zeadally}, {and} \bibinfo{person}{Jonathan Cochran}.}
  \bibinfo{year}{2018}\natexlab{}.
\newblock \showarticletitle{A review and an empirical analysis of privacy
  policy and notices for consumer Internet of things}.
\newblock \bibinfo{journal}{\emph{Security and Privacy}} \bibinfo{volume}{1},
  \bibinfo{number}{3} (\bibinfo{year}{2018}), \bibinfo{pages}{e15}.
\newblock


\bibitem[Qi et~al\mbox{.}(2020)]%
        {qi2020stanza}
\bibfield{author}{\bibinfo{person}{Peng Qi}, \bibinfo{person}{Yuhao Zhang},
  \bibinfo{person}{Yuhui Zhang}, \bibinfo{person}{Jason Bolton}, {and}
  \bibinfo{person}{Christopher~D. Manning}.} \bibinfo{year}{2020}\natexlab{}.
\newblock \showarticletitle{Stanza: A {Python} Natural Language Processing
  Toolkit for Many Human Languages}. In \bibinfo{booktitle}{\emph{Proceedings
  of the 58th Annual Meeting of the Association for Computational Linguistics:
  System Demonstrations}}.
\newblock
\urldef\tempurl%
\url{https://nlp.stanford.edu/pubs/qi2020stanza.pdf}
\showURL{%
\tempurl}


\bibitem[Rahmouni et~al\mbox{.}(2014)]%
        {rahmouni2014semantic}
\bibfield{author}{\bibinfo{person}{Hanene~Boussi Rahmouni},
  \bibinfo{person}{Kamran Munir}, \bibinfo{person}{Marco~Casassa Mont}, {and}
  \bibinfo{person}{Tony Solomonides}.} \bibinfo{year}{2014}\natexlab{}.
\newblock \showarticletitle{Semantic generation of clouds privacy policies}. In
  \bibinfo{booktitle}{\emph{International Conference on Cloud Computing and
  Services Science}}. Springer, \bibinfo{pages}{15--30}.
\newblock


\bibitem[Schwenk and Koehn(2008)]%
        {schwenk2008large}
\bibfield{author}{\bibinfo{person}{Holger Schwenk} {and}
  \bibinfo{person}{Philipp Koehn}.} \bibinfo{year}{2008}\natexlab{}.
\newblock \showarticletitle{Large and diverse language models for statistical
  machine translation}. In \bibinfo{booktitle}{\emph{Proceedings of the Third
  International Joint Conference on Natural Language Processing: Volume-II}}.
\newblock


\bibitem[Shi et~al\mbox{.}(2022)]%
        {shi2022compressing}
\bibfield{author}{\bibinfo{person}{Jieke Shi}, \bibinfo{person}{Zhou Yang},
  \bibinfo{person}{Bowen Xu}, \bibinfo{person}{Hong~Jin Kang}, {and}
  \bibinfo{person}{David Lo}.} \bibinfo{year}{2022}\natexlab{}.
\newblock \showarticletitle{Compressing Pre-trained Models of Code into 3 MB}.
  In \bibinfo{booktitle}{\emph{37th IEEE/ACM International Conference on
  Automated Software Engineering}}. \bibinfo{pages}{1--12}.
\newblock


\bibitem[Spiekermann et~al\mbox{.}(2001)]%
        {spiekermann2001privacy}
\bibfield{author}{\bibinfo{person}{Sarah Spiekermann}, \bibinfo{person}{Jens
  Grossklags}, {and} \bibinfo{person}{Bettina Berendt}.}
  \bibinfo{year}{2001}\natexlab{}.
\newblock \showarticletitle{E-privacy in 2nd generation E-commerce: privacy
  preferences versus actual behavior}. In \bibinfo{booktitle}{\emph{Proceedings
  of the 3rd ACM conference on Electronic Commerce}}. \bibinfo{pages}{38--47}.
\newblock


\bibitem[Tang et~al\mbox{.}(2021)]%
        {tang2021defining}
\bibfield{author}{\bibinfo{person}{Jenny Tang}, \bibinfo{person}{Hannah
  Shoemaker}, \bibinfo{person}{Ada Lerner}, {and} \bibinfo{person}{Eleanor
  Birrell}.} \bibinfo{year}{2021}\natexlab{}.
\newblock \showarticletitle{Defining privacy: How users interpret technical
  terms in privacy policies}.
\newblock \bibinfo{journal}{\emph{Proceedings on Privacy Enhancing
  Technologies}} \bibinfo{volume}{2021}, \bibinfo{number}{3}
  (\bibinfo{year}{2021}).
\newblock


\bibitem[Tao et~al\mbox{.}(2022)]%
        {tao2022compression}
\bibfield{author}{\bibinfo{person}{Chaofan Tao}, \bibinfo{person}{Lu Hou},
  \bibinfo{person}{Wei Zhang}, \bibinfo{person}{Lifeng Shang},
  \bibinfo{person}{Xin Jiang}, \bibinfo{person}{Qun Liu}, \bibinfo{person}{Ping
  Luo}, {and} \bibinfo{person}{Ngai Wong}.} \bibinfo{year}{2022}\natexlab{}.
\newblock \showarticletitle{Compression of generative pre-trained language
  models via quantization}.
\newblock \bibinfo{journal}{\emph{arXiv preprint arXiv:2203.10705}}
  (\bibinfo{year}{2022}).
\newblock


\bibitem[Tinggi et~al\mbox{.}(2011)]%
        {tinggi2011customers}
\bibfield{author}{\bibinfo{person}{Michael Tinggi}, \bibinfo{person}{Shaharudin
  Jakpar}, \bibinfo{person}{Ting~Bick Chin}, {and} \bibinfo{person}{Junaid~M
  Shaikh}.} \bibinfo{year}{2011}\natexlab{}.
\newblock \showarticletitle{Customers? Confidence and trust towards privacy
  policy: a conceptual research of hotel revenue management}.
\newblock \bibinfo{journal}{\emph{International Journal of Revenue Management}}
  \bibinfo{volume}{5}, \bibinfo{number}{4} (\bibinfo{year}{2011}),
  \bibinfo{pages}{350--368}.
\newblock


\bibitem[Toch et~al\mbox{.}(2010)]%
        {toch2010generating}
\bibfield{author}{\bibinfo{person}{Eran Toch}, \bibinfo{person}{Norman~M
  Sadeh}, {and} \bibinfo{person}{Jason Hong}.} \bibinfo{year}{2010}\natexlab{}.
\newblock \showarticletitle{Generating default privacy policies for online
  social networks}.
\newblock In \bibinfo{booktitle}{\emph{CHI'10 Extended Abstracts on Human
  Factors in Computing Systems}}. \bibinfo{pages}{4243--4248}.
\newblock


\bibitem[Touvron et~al\mbox{.}(2023)]%
        {touvron2023llama}
\bibfield{author}{\bibinfo{person}{Hugo Touvron}, \bibinfo{person}{Thibaut
  Lavril}, \bibinfo{person}{Gautier Izacard}, \bibinfo{person}{Xavier
  Martinet}, \bibinfo{person}{Marie-Anne Lachaux}, \bibinfo{person}{Timothée
  Lacroix}, \bibinfo{person}{Baptiste Rozière}, \bibinfo{person}{Naman Goyal},
  \bibinfo{person}{Eric Hambro}, \bibinfo{person}{Faisal Azhar},
  \bibinfo{person}{Aurelien Rodriguez}, \bibinfo{person}{Armand Joulin},
  \bibinfo{person}{Edouard Grave}, {and} \bibinfo{person}{Guillaume Lample}.}
  \bibinfo{year}{2023}\natexlab{}.
\newblock \bibinfo{title}{LLaMA: Open and Efficient Foundation Language
  Models}.
\newblock
\newblock
\showeprint[arxiv]{2302.13971}~[cs.CL]


\bibitem[Ullah et~al\mbox{.}(2020)]%
        {ullah2020privacy}
\bibfield{author}{\bibinfo{person}{Imdad Ullah}, \bibinfo{person}{Roksana
  Boreli}, {and} \bibinfo{person}{Salil~S Kanhere}.}
  \bibinfo{year}{2020}\natexlab{}.
\newblock \showarticletitle{Privacy in targeted advertising: A survey}.
\newblock \bibinfo{journal}{\emph{arXiv preprint arXiv:2009.06861}}
  (\bibinfo{year}{2020}).
\newblock


\bibitem[Wilson et~al\mbox{.}(2016)]%
        {wilson2016creation}
\bibfield{author}{\bibinfo{person}{Shomir Wilson}, \bibinfo{person}{Florian
  Schaub}, \bibinfo{person}{Aswarth~Abhilash Dara}, \bibinfo{person}{Frederick
  Liu}, \bibinfo{person}{Sushain Cherivirala}, \bibinfo{person}{Pedro~Giovanni
  Leon}, \bibinfo{person}{Mads~Schaarup Andersen}, \bibinfo{person}{Sebastian
  Zimmeck}, \bibinfo{person}{Kanthashree~Mysore Sathyendra},
  \bibinfo{person}{N~Cameron Russell}, {et~al\mbox{.}}}
  \bibinfo{year}{2016}\natexlab{}.
\newblock \showarticletitle{The creation and analysis of a website privacy
  policy corpus}. In \bibinfo{booktitle}{\emph{Proceedings of the 54th Annual
  Meeting of the Association for Computational Linguistics (Volume 1: Long
  Papers)}}. \bibinfo{pages}{1330--1340}.
\newblock


\bibitem[Windl et~al\mbox{.}(2022)]%
        {windl2022automating}
\bibfield{author}{\bibinfo{person}{Maximiliane Windl}, \bibinfo{person}{Niels
  Henze}, \bibinfo{person}{Albrecht Schmidt}, {and}
  \bibinfo{person}{Sebastian~S Feger}.} \bibinfo{year}{2022}\natexlab{}.
\newblock \showarticletitle{Automating Contextual Privacy Policies: Design and
  Evaluation of a Production Tool for Digital Consumer Privacy Awareness}. In
  \bibinfo{booktitle}{\emph{CHI Conference on Human Factors in Computing
  Systems}}. \bibinfo{pages}{1--18}.
\newblock


\bibitem[Xie and Karan(2019)]%
        {xie2019consumers}
\bibfield{author}{\bibinfo{person}{Wenjing Xie} {and} \bibinfo{person}{Kavita
  Karan}.} \bibinfo{year}{2019}\natexlab{}.
\newblock \showarticletitle{Consumers’ privacy concern and privacy protection
  on social network sites in the era of big data: empirical evidence from
  college students}.
\newblock \bibinfo{journal}{\emph{Journal of Interactive Advertising}}
  \bibinfo{volume}{19}, \bibinfo{number}{3} (\bibinfo{year}{2019}),
  \bibinfo{pages}{187--201}.
\newblock


\bibitem[Yang(2023)]%
        {why2023yang}
\bibfield{author}{\bibinfo{person}{Jingfeng Yang}.}
  \bibinfo{year}{2023}\natexlab{}.
\newblock \bibinfo{booktitle}{\emph{Why did all of the public reproduction of
  GPT-3 fail? In which tasks should we use GPT-3.5/ChatGPT?}}
\newblock
\urldef\tempurl%
\url{https://jingfengyang.github.io/gpt}
\showURL{%
\tempurl}
\newblock
\shownote{Accessed: 2022-03-27}.


\bibitem[Yu et~al\mbox{.}(2018)]%
        {TSE_2018}
\bibfield{author}{\bibinfo{person}{Le Yu}, \bibinfo{person}{Xiapu Luo},
  \bibinfo{person}{Jiachi Chen}, \bibinfo{person}{Hao Zhou},
  \bibinfo{person}{Tao Zhang}, \bibinfo{person}{Henry Chang}, {and}
  \bibinfo{person}{Hareton~KN Leung}.} \bibinfo{year}{2018}\natexlab{}.
\newblock \showarticletitle{Ppchecker: Towards accessing the trustworthiness of
  android apps’ privacy policies}.
\newblock \bibinfo{journal}{\emph{IEEE Transactions on Software Engineering}}
  \bibinfo{volume}{47}, \bibinfo{number}{2} (\bibinfo{year}{2018}),
  \bibinfo{pages}{221--242}.
\newblock


\bibitem[Yu et~al\mbox{.}(2015)]%
        {yu2015autoppg}
\bibfield{author}{\bibinfo{person}{Le Yu}, \bibinfo{person}{Tao Zhang},
  \bibinfo{person}{Xiapu Luo}, {and} \bibinfo{person}{Lei Xue}.}
  \bibinfo{year}{2015}\natexlab{}.
\newblock \showarticletitle{Autoppg: Towards automatic generation of privacy
  policy for android applications}. In \bibinfo{booktitle}{\emph{Proceedings of
  the 5th Annual ACM CCS Workshop on Security and Privacy in Smartphones and
  Mobile Devices}}. \bibinfo{pages}{39--50}.
\newblock


\bibitem[Zhang et~al\mbox{.}(2022)]%
        {zhang2022usable}
\bibfield{author}{\bibinfo{person}{Shikun Zhang}, \bibinfo{person}{Yuanyuan
  Feng}, \bibinfo{person}{Yaxing Yao}, \bibinfo{person}{Lorrie~Faith Cranor},
  {and} \bibinfo{person}{Norman Sadeh}.} \bibinfo{year}{2022}\natexlab{}.
\newblock \showarticletitle{How Usable Are iOS App Privacy Labels?}
\newblock \bibinfo{journal}{\emph{UMBC Faculty Collection}}
  (\bibinfo{year}{2022}).
\newblock


\bibitem[Zimmeck et~al\mbox{.}(2021)]%
        {zimmeck2021privacyflash}
\bibfield{author}{\bibinfo{person}{Sebastian Zimmeck}, \bibinfo{person}{Rafael
  Goldstein}, {and} \bibinfo{person}{David Baraka}.}
  \bibinfo{year}{2021}\natexlab{}.
\newblock \showarticletitle{PrivacyFlash Pro: Automating Privacy Policy
  Generation for Mobile Apps.}. In \bibinfo{booktitle}{\emph{NDSS}}.
\newblock


\bibitem[Zimmeck et~al\mbox{.}(2019)]%
        {zimmeck2019maps}
\bibfield{author}{\bibinfo{person}{Sebastian Zimmeck}, \bibinfo{person}{Peter
  Story}, \bibinfo{person}{Daniel Smullen}, \bibinfo{person}{Abhilasha
  Ravichander}, \bibinfo{person}{Ziqi Wang}, \bibinfo{person}{Joel~R
  Reidenberg}, \bibinfo{person}{N~Cameron Russell}, {and}
  \bibinfo{person}{Norman Sadeh}.} \bibinfo{year}{2019}\natexlab{}.
\newblock \showarticletitle{Maps: Scaling privacy compliance analysis to a
  million apps}.
\newblock \bibinfo{journal}{\emph{Proc. Priv. Enhancing Tech.}}
  \bibinfo{volume}{2019} (\bibinfo{year}{2019}), \bibinfo{pages}{66}.
\newblock


\end{thebibliography}

\end{document}